\documentclass[useAMS,usenatbib]{aa}
\usepackage{deluxetable}
\usepackage{natbib}
\usepackage{txfonts}
\usepackage{graphicx}
\usepackage{longtable}
\usepackage{txfonts}
\usepackage{rotating}
\usepackage{epsf}
\usepackage{psfrag}
\usepackage{amssymb}
\usepackage{url}
\bibpunct[,]{}{}{;}{a}{}{,}
\def\sec{\ifmmode {}^{\prime\prime}\else ${}^{\prime\prime}$\fi~}

\def\magdot{\ifmmode {}^{\rm m}\!\!\!.\, \else ${}^{\rm m}\!\!\!.\,$\fi}
\def\daydot{\ifmmode {}^{\rm d}\!\!\!.\, \else ${}^{\rm d}\!\!\!.\,$\fi}
\def\asec{\ifmmode ^{\prime\prime}\else$^{\prime\prime}$\fi}

\begin{document} 

\title{The Nainital-Cape Survey-III : A Search for  Pulsational Variability in Chemically Peculiar Stars\thanks{Table \ref{null}, Figure \ref{unpre01} and Figure \ref{pre01} are only available in  electronic form at http://www.edpsciences.org}}

\author{S. Joshi \inst{1}, D. L. Mary \inst{2}, N. K. Chakradhari \inst{3}, S. K. Tiwari \inst{1}, C. Billaud \inst{2}
}

\offprints{Santosh Joshi}

\authorrunning{S. Joshi et al.}

\institute{
Aryabhatta Research Institute of Observational Sciences (ARIES), Manora Peak, Nainital, India\\
\email{santosh@aries.res.in}
\and
Universit\'e de Nice Sophia Antipolis, Observatoire de la C\^ote d'Azur, Nice, France \\
\email{david.mary@unice.fr,camille.billaud@wanadoo.fr}
\and
School of Studies in Physics and Astrophysics, Pt. Ravishankar Shukla University, Raipur, India \\
\email{nkchakradhari@gmail.com}
}

\date{Received; accepted}

\authorrunning{S. Joshi et al.}
\titlerunning{The Nainital-Cape Survey-III}

\abstract
{The Nainital-Cape survey is a dedicated research programme to search and study pulsational variability in chemically peculiar stars in the Northern Hemisphere.}
{The aim of the survey is to search such chemically peculiar stars which are pulsationally unstable.}
{The observations of the sample stars were carried out in high-speed photometric mode using a three-channel fast photometer attached to the 1.04-m Sampurnanand telescope at ARIES.}
{The new photometric observations confirmed that the pulsational period of star HD\,25515 is 2.78-hrs. The repeated time-series observations of HD\,113878 and HD\,118660 revealed that previously known frequencies are indeed present in the new data sets. We have estimated the distances, absolute magnitudes, effective temperatures and luminosities of these stars. Their positions in the H-R diagram indicate that HD\,25515 and HD\,118660 lie near the main-sequence while HD\,113878 is an evolved star. We also present a catalogue of 61 stars classified as null results, along with the corresponding 87 frequency spectra taken over the time scale  2002-2008.  A  statistical analysis of these null results shows, by comparison with past data, that the power of the noise in the light curves has slightly increased during the last few years.}
{}
   

\maketitle
 
\section{Introduction}
\label{introduction}
\label{intro}
A survey project called ``Nainital-Cape Survey'' was initiated in 1997 between the Aryabhatta Research Institute of Observational Sciences (ARIES), Nainital, and the South African Astronomical
Observatory (SAAO), Capetown, to search and study the  pulsational variability in chemically peculiar (CP) A-F stars.
The atmospheres of these stars have an inhomogeneous distribution of chemical elements which 
are frequently seen as abundance spots on stellar surfaces and as clouds of chemical elements concentrated at
certain depths along the stellar radii (see, for example, Kochukhov et al. \cite{kochukhov04a,kochukhov04b}; Lueftinger et al. \cite{lueftinger03}; Ryabchikova et al. \cite{ryabchikova06,ryabchikova05, ryabchikova02}).  The inhomogeneity with high abundance gradients is explained by microscopic radiative diffusion processes depending on the balance between gravitational pull and uplift by the radiation field through absorption in spectral lines (Michaud \cite{michaud70}, Michaud et al. \cite{michaud81};  LeBlanc \& Monin \cite{leblanc09}). Therefore, the CP stars are excellent test objects for astrophysical processes like diffusion, convection, and stratification in stellar atmospheres in the presence of rather strong magnetic fields.

  A class of CP (CP2) stars known as Ap (A-peculiar) stars, which rotate considerably more slowly than spectroscopically normal main sequence stars have similar effective temperatures, strong magnetic fields , normal or below-solar concentrations of light and iron-peak elements, and overabundances of rare Earth elements.  A fraction of cool Ap stars in the effective temperature range of $\approx$ 6400 to 8100 K (Kochukhov \& Bagnulo \cite{kochukhov06}) exhibit anomalously strong Sr, Cr and Eu lines coupled with high-magnetic fields from few to as high as 24.5 kGauss (Kurtz et al. \cite{kurtz06b}). Those stars which pulsate  in the period of range 5.65 to 22-min (Kreidl \cite{kreidl85}, Kochukhov \cite{kochukhov08a}), with photometric amplitude variations less than 10 milli-magnitude (mmag) in Johnson B filter, and spectroscopic radial velocity variations of 0.05 to 5-kms$^{-1}$ are  termed  rapidly oscillating Ap (roAp) stars. The roAp pulsations are attributed to low-order ($l<$ 3), high-overtones ($n \gg l$), non-radial $p$-modes and are classically explained by the oblique pulsator model (Kurtz \cite{kurtz82}). This model supposes the pulsation axis is aligned to the magnetic axis which is oblique to the rotation axis. As the star rotates the variations in the  pulsational amplitude can be observed by a distant observer. Recent theoretical studies of roAp stars show that the axis of the $p$-mode pulsations is not exactly aligned with the magnetic field because of the centrifugal force (Bigot \& Dziembowski \cite{bigot02}).  Excitation of pulsations in  roAp stars  is governed by $\kappa$-mechanism acting in the H \textsc{I} ionization zone, with the additional influence from the magnetic quenching of convection and composition gradients built up by  the atomic diffusion (Balmoforth et al. \cite{balmforth01}; Cunha \cite{cunha02}; Vauclair \& Th\'eado \cite{vauclair04}). These magneto-acoustic pulsators provide a unique opportunity to study the  pulsations in the presence of chemical inhomogeneities and strong magnetic fields. 

Apart from the roAp stars,  another class of CP stars (CP1) known as Am (A-metallic) stars  and preferably found within close binary systems, exhibit low-rotational velocities, lack of magnetic fields, apparent underabundance of calcium and scandium, and overabundance of Fe-peak elements. The former near exclusion between pulsational variability and  Am-type chemical peculiarity has evolved considerably in the recent years,  and this dichotomy is no longer a clear-cut. Indeed, pulsations of the $\delta$-Scuti type\footnote{$\delta$-Scuti stars are pulsating variables with spectral type from A2 to F0 and luminosities ranging from zero-age main sequence to about two magnitude above the main sequence. Their known periods range from 18.12-min (HD\,34282, Amado et al. \cite{amado04}) to about 8-hrs, with amplitude of light variations up to a few tens of mmag.  The pulsations in $\delta$-Scuti stars are characterized by low-order, low-degree radial and/or non-radial modes which are governed by $\kappa$-mechanism operated in the He \textsc{II} ionization zone.} have been evidenced in many Am stars (e.g. Kurtz \cite{kurtz89}; Martinez et al. \cite{martinez99}; Kurtz \& Martinez \cite{kurtz00}; Joshi et al. \cite{joshi03}, Gonz\'alez et al. \cite{gonz08}), but our understanding of the physics of pulsating Am stars is still far from being complete. Theoretical models can for instance  partially explain that evolved Am stars can pulsate ($\kappa$-mechanism in He \textsc{II} can drive the pulsations in evolved Am stars) and that marginal Am stars can be low-amplitude $\delta$-Scuti stars (a sufficient amount of He remains to drive low-amplitude oscillations). However,  less well  understood are the mechanisms of pulsations in classical Am stars (Kurtz \& Martinez \cite{kurtz00}), $\delta$-Scuti type pulsations in Ap stars (Gonz\'alez et al. \cite{gonz08}), origin of  the abundance anomalies either from a superficial or from a much deeper mixing zone (LeBlanc et al. \cite{leblanc08}, \cite{leblanc09}),  co-existence of $\delta$-Scuti and  $\gamma$-Doradus type pulsations in Am stars (King et al. \cite{king06}; Henry \& Fekel \cite{henry05}), limits where chemical peculiarity and pulsations can coexist and where they are mutually exclusive (Hekker et al. \cite{hekker08}) and possible excitation of solar like oscillations in Am stars (Carrier et al. \cite{carrier07}). 

The rich pulsation spectra of roAp and pulsating Am stars does not only help us to study  the interaction of pulsations and chemical peculiarities : several important astrophysical parameters such as   mass, luminosity, rotational period  or magnetic field strength can also be inferred, or at least constrained (Balmforth et al. \cite{balmforth01}; Cunha \cite{cunha02}; Saio \cite{saio05}; Bruntt et al. \cite{bruntt09}). 

However, despite  several searches (e.g. Nelson \& Kreidl \cite{nelson93}; Martinez \& Kurtz \cite{martinez94}; Handler \& Paunzen \cite{handler99}; Martinez et al. \cite{martinez01}; Dorokhova \& Dorokhov \cite{dorokhova05}; Joshi et al. \cite{joshi06}) only about 40 roAp stars are known to  date (Kochukhov \cite{kochukhov08a}). This dearth of roAp stars demands high-precision systematic surveys. In this context, the ``Nainital-Cape'' survey is an effort to search pulsational variability in Ap and Am stars.   In the last decade, several results were obtained from this survey. They  are summarized in Table \ref{surveyresult} : HD\,12098 was discovered as a roAp star (Girish et al. \cite{girish01}), and pulsations of the $\delta$-Scuti type  were discovered in 6 CP stars : HD\,13038, HD\,13079 (Martinez et al. \cite{martinez01}), HD\,98851, HD\,102480 (Joshi et al. \cite{joshi03}), HD\,113878 and HD\,118660 (Joshi et al. \cite{joshi06}).  

\begin{table}
\begin{center}
\label{surveyresult}
\caption{: Pulsating variables discovered during the ``Nainital-Cape survey''.  HD\,98851 and HD\,102480 are unusual pulsators in  the sense they exhibit  alternating high- and low-maxima (i.e., the amplitude of the pulsations is being high and low in a cyclic way).}
\bigskip 
\begin{scriptsize}
\begin{tabular}{|r|r|r|c|}
\hline
Star&$P_1$&$P_2$&Comments \\
 HD& (min) & (min) & \\
\hline
12098 & 7.6 & - & roAp type \\
13038 & 28.0 & 34.0 & Multi-periodic $\delta$-Scuti type  \\
13079 & 73.2 & - & $\delta$-Scuti type  \\
98851 & 81.0 & 162.0 & Alternating High and Low-maxima \\
102480 & 156.0 & 84.0 & Alternating High and Low-maxima\\
113878 & 138.6 & - & $\delta$-Scuti type  \\
118660 & 60.0 & 151.2 & Multi-periodic $\delta$-Scuti type  \\
\hline
\end{tabular}
\end{scriptsize}
\end{center}
\end{table}

Since the 1980's, the high-speed photometric technique is being used to search and study  the photometric light variations in  short period ($\sim$ min) pulsating variables. In parallel,  high-resolution spectroscopic techniques were recently used to measure the corresponding radial velocity (RV) variations and yielded many interesting results (e.g. Kochukhov \& Ryabchikova \cite{kochukhov01}; Mkrtichian et al. \cite{mkrtichian03}, Hatzes \& Mkrtichian \cite{hatzes04}; Elkin et al. \cite{elkin05}; Kurtz et al. \cite{kurtz06a}; Ryabchikova et al. \cite{ryabchikova07}; Kochukhov et al. \cite{kochukhov08b}; Gonz\'alez et al. \cite{gonz08}).  The RV measurements are a very useful technique towards the derivation of  horizontal informations  obtained from the line profile variability. Vertical information are extracted from the differential analysis of both amplitude and phase of spectral lines formed at different atmospheric heights. Also the phase information of individual spectral lines and oscillation modes is an important tool to constrain the chemical stratification and the physical models of the atmosphere. For these reasons, RV studies of individual lines provide a very powerful tool towards an atmospheric tomography of  roAp stars.

If spectroscopic studies are such an efficient tool, why to continue doing photometric surveys ? Despite the improved sensitivity of the high-resolution spectroscopic monitoring  for searching low-amplitude oscillations in roAp candidates, the major drawback of this technique remains the small amount of observing time available at large telescopes. Traditional photometric  surveys are performed on small telescopes, and are consequently relatively cheap (both in cost and telescope time). For instance, Kurtz discovered the first roAp star HD\,101065  by means of high-speed photometry using the 50-cm telescope of SAAO (Kurtz \cite{kurtz78}). Indeed, this argument in favor of photometry assumes that the photometric instruments are periodically upgraded : since the 1970s, most or all bright stars that are observable with good signal to noise ratios using small telescopes and traditional detectors have been checked for pulsation at the mmag level. In order to push our investigations towards fainter objects,  the technology of photometric surveys must consequently be updated with state-of-the-art instruments. This is one reason why  ARIES decided to give a new impulsion to the survey by installing bigger telescopes at a better astronomical site (see Sec. 5.2, and Sagar et al. \cite{sagar00}; Stalin et al. \cite{stalin01}).

This paper is the third of a series (Paper I : Martinez et al. \cite{martinez01}, Paper II : Joshi et al. \cite{joshi06}). It is organized as follows : the selection criteria are briefly described in Sec. \ref{selection} and  the technique of the observation and data analysis is described in Sec. \ref{observations}. Results from the new observations  of HD\,25515, HD\,113878 and HD\,118660, along with some astrophysical parameters of these stars are presented in Sec. \ref{newobservations}. Sec. \ref{nullresults} presents the null results and their analysis. Finally some conclusions are drawn in Sec. \ref{conclusions}.

\section{Selection of the Candidates}
\label{selection}

To the date, there are no firmly established selection criteria to maximize the chances of detecting new variables.
Therefore our primary criterion was to choose  candidates presenting Str\"{o}mgren photometric indices similar to those  Ap and Am stars which exhibit pulsational variability (Hauck \& Mermilliod \cite{hauck98}).  For this, the following range of Str\"{o}mgren photometric indices was used to select the samples: $0.46 \le c_1 \le 0.88$; $0.19 \le m_1 \le 0.33$; $2.69 \le \beta \le 2.88$; $0.08 \leq b-y \le 0.31$; $-0.12 \le \delta m_1 \le 0.02$ and $\delta c_1 \le 0.04$, where $c_1$ is the Balmer discontinuity parameter (luminosity indicator), $m_1$ is the line-blanketing parameter (metallicity indicator), $\beta$ is the $H_{\beta}$ line strength index  (indicator of temperature in the spectral range A3 to F2 and reasonably free from reddening), and $b-y$ is also an indicator of temperature, affected by reddening. A more negative value of $\delta m_1$ [$m_1$(standard)-$m_1$(observed)] and $\delta c_1$ [$c_1$(observed)-$c_1$(standard)]  means that the star is more peculiar  (Crawford \cite{crawford75}, \cite{crawford79}). However, we have slightly extended (by $\approx$10\%) the corresponding range of indices to include the evolved and cooler stars, hence there are sample stars not falling  in the range of Str\"{o}mgren photometric indices mentioned above.  The candidates were also selected  from the general catalogue of Ap and Am stars (Renson et al. \cite{renson91}) and the fifth Michigan catalogue (Houk \& Swift \cite{houk99}). The roAp stars are magnetic and of the SrCrEu type,  and pulsating Am stars are non-magnetic. Since both belong to the spectral class of late A to early F,  samples  were also selected  from the  Kudryavtsev et al. (\cite{kudryavtsev06}); Bychkov et al. (\cite{bychkov03}) and H$\o$g et al. (\cite{hog00}). 

\section{Observations and Data Analysis}
\label{observations}

The time-series photometric data consist of continuous $10$-sec integrations obtained through a Johnson $B$-filter. An aperture of 30$^{\prime\prime}$ was used to minimize flux variations caused by seeing fluctuations and guiding errors. The data reduction process comprises of visual
inspection of the light curve to identify and remove the bad data
points, correction for coincident counting losses, subtraction of
the interpolated sky background and correction for the mean atmospheric extinction.
After applying these corrections, the times of the mid-points of  each data samples are converted
into Heliocentric Julian dates (HJD) with an accuracy of $10^{-5}$ day
($\approx$1 s). The reduced data comprise a time-series of HJD and B-magnitude with respect to the mean of the light curve. 

To search periodic signals,  the time-series data  were classically analyzed using a fast algorithm based on Deeming's Discrete Fourier Transform (DFT) for unequally spaced data (Deeming \cite{deeming75}). The first step of the analysis is to inspect the DFTs of each individual light curve where the dominant frequency $f_1$ is identified. To identify other frequencies present in the data, a sinusoid corresponding to the dominant frequency, amplitude and phase (of the form $A_1 \cos(2\pi f_1t+\phi_1))$ is subtracted from the time-series. The residuals of this fit are then used to compute the DFT again, the resulting dominant frequency is identified as $f_2$ and so on.  This prewhitening procedure was repeated until the residuals were judged to be at the noise level. To search additional frequencies and to define the known frequencies in a better way, we also computed  the DFTs of the combined data sets for each star.  We comment further on the data analysis in the next section.

\section{New Observations of HD\,25515, HD\,113878 and HD\,118660}
\label{newobservations}
The following subsections describe the results from the follow-up observations and discuss some of the fundamental astrophysical parameters of the individual stars.

\subsection{HD\,25515}
\label{hd25515}
\begin{figure*}[ht]
\hbox{
\includegraphics[width=9.5cm,height=7cm]{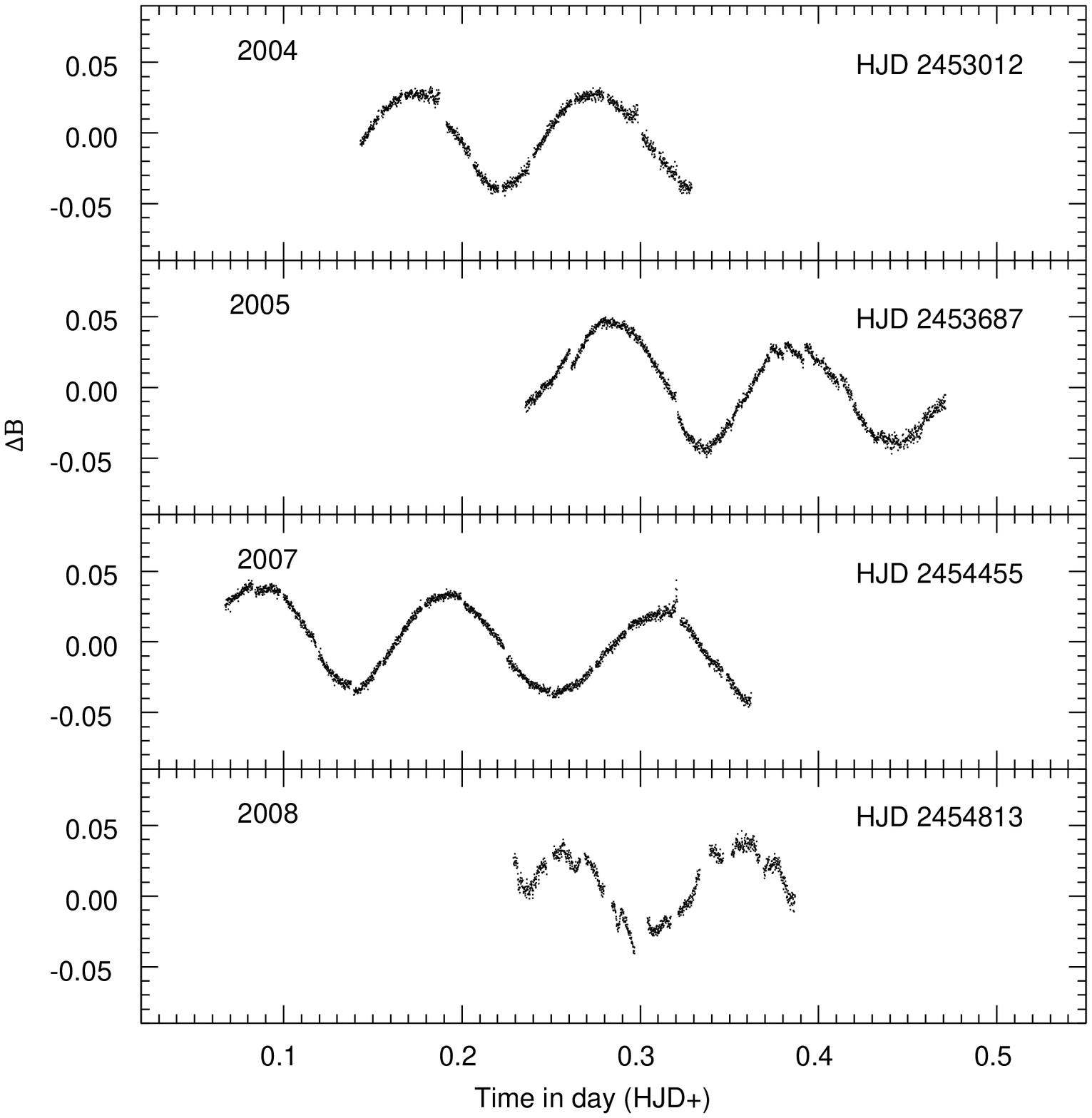}
\includegraphics[width=9.5cm,height=7cm]{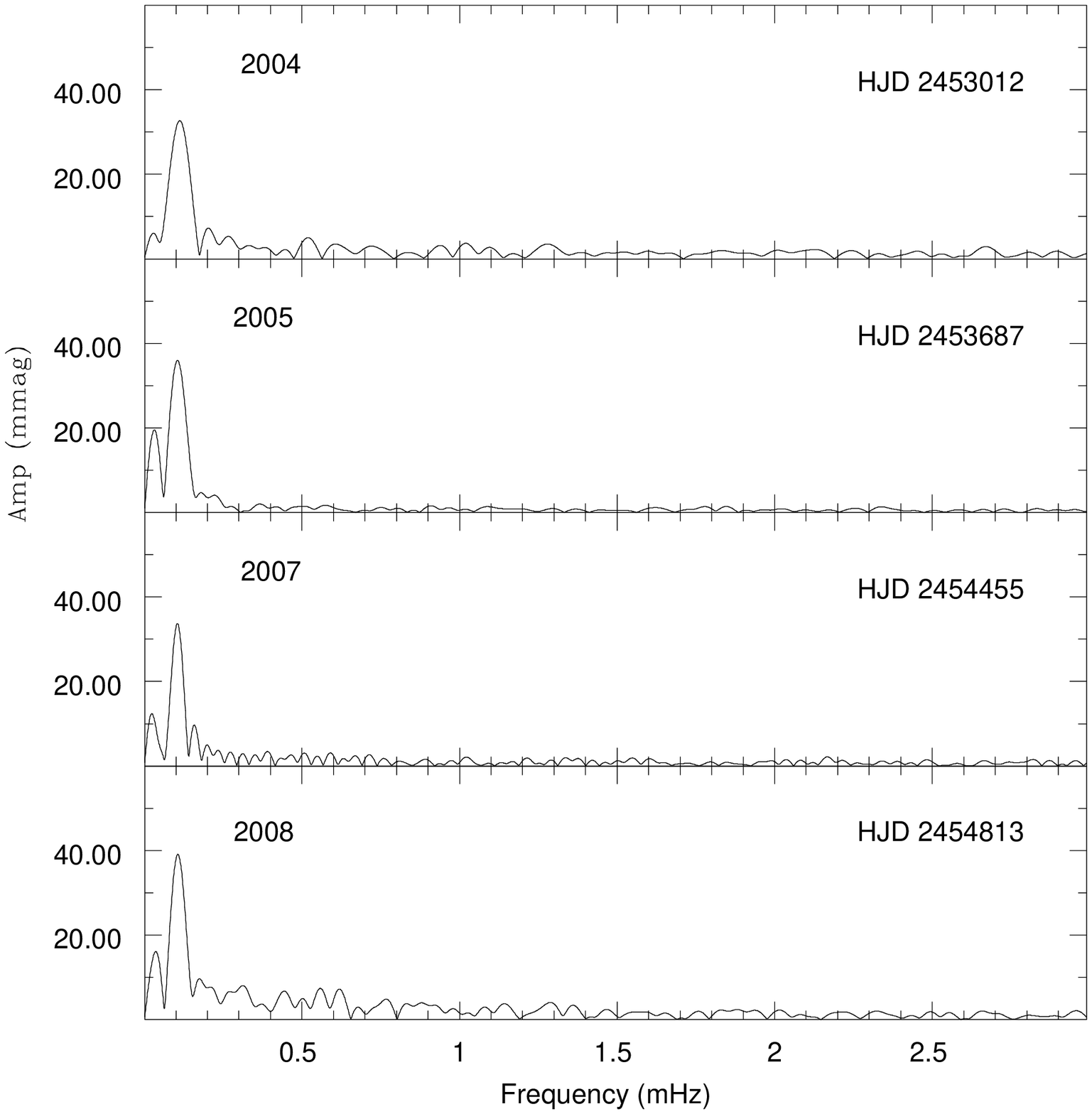}
}
\caption{: Four typical light curves (left) and their corresponding amplitude spectra (right) of HD\,25515 obtained on four different observing seasons.}
\label{hd25515lc}
\end{figure*}

The Str\"{o}mgren indices of HD\,25515 are $b-y$ = 0.262, $m_1$ = 0.177, $c_1$ = 0.745 and $H_{\beta}$ = 2.706 (Hauck \& Mermilliod \cite{hauck98}).
On the basis of these peculiar indices, HD\,25515 was included in the survey programme. 
Chaubey \& Kumar (\cite{chaubey05}) announced the discovery of a pulsational period of 2.8-hrs in HD\,25515, but so far no light curves had been published in the literature.
Since year 2004, a total of 57.52-hrs data was collected on 13 nights. The left panel of Fig. \ref{hd25515lc} shows four typical light curves of HD\,25515 and the corresponding amplitude spectra of these light curves are shown in the right panel. 
\begin{table}
\begin{center}
\caption{: Journal of observations of HD\,25515. A value $x$ in the Start Time HJD columns means Heliocentric Julian Date 2450000+$x$. The data sets marked with asterisks are used for frequency analysis to search for the additional frequencies and define the known frequencies in a better way.}
\medskip
\begin{scriptsize}
\begin{tabular}{|r|c|c|c|c|r|}
\hline
S. No. & UT & Start time & $\Delta$t  & $f_1,f_2$& $A_1,A_2$~~~\\
    &Date & HJD & (hr) & (mHz)~~~ &(mmag)\\
\hline
1. & 06-01-04 &3011.14446 & 1.54 & 0.12 $\pm$ 0.08 & 21.10 \\
      &          &           &      & 0.14 $\pm$ 0.08 & 5.00  \\          
2. & 07-01-04 & 3012.14302$^*$ & 3.94 & 0.11 $\pm $ 0.04 & 32.68 \\
      &          &           &      & 0.19 $\pm$ 0.05& 4.15\\      
3. & 11-01-04 & 3016.13736$^*$ & 3.96      & 0.10 $\pm$ 0.03& 17.39 \\
      &          &           &      & 0.18 $\pm$ 0.02& 2.51\\      
4. & 12-01-04 & 3017.18221 & 2.21 & 0.13 $\pm$ 0.07 & 28.78 \\
      &          &           &      & 0.21 $\pm$ 0.05& 9.76\\         
5. & 05-02-04 & 3041.08654$^*$ & 2.47 & 0.11 $\pm$ 0.07 & 15.07 \\
      &          &      &      &  0.16 $\pm$ 0.09 & 2.29 \\               
6. & 10-02-04 & 3046.07930$^*$ & 1.92 & 0.12 $\pm$ 0.07 & 15.91 \\
      &          &       &      & 0.27 $\pm$ 0.08 & 4.70 \\
7. & 16-10-05 & 3660.27181 & 5.07 & 0.10 $\pm$ 0.03 & 29.76 \\
      &          &       &      & 0.14 $\pm$ 0.03 & 10.16 \\         
8. & 17-11-05 & 3661.32755 & 3.86 & 0.11 $\pm$ 0.03 & 31.51 \\
      &          &      &       & 0.04 $\pm$ 0.03 & 12.92 \\            
9. & 12-11-05 & 3687.23566$^*$ & 5.40 & 0.10 $\pm$ 0.03 & 35.99 \\
      &          &      &       & 0.03 $\pm$ 0.02 & 11.29 \\  
10. & 13-11-05 & 3688.30109 & 4.24 & 0.04 $\pm$ 0.03 & 46.88 \\
      &          &      &       & 0.06 $\pm$ 0.03 & 13.37 \\    
11. & 14-11-05 & 3689.14266$^*$ & 5.48 & 0.03 $\pm$ 0.02 & 78.01 \\
      &          &      &       & 0.12 $\pm$ 0.03 & 25.03 \\                 
12. & 20-12-07 & 4455.06725 & 6.59 & 0.10 $\pm$ 0.02 & 33.73 \\
      &          &      &       & 0.03 $\pm$ 0.02 & 9.33 \\
13. & 12-12-08 & 4813.14728$^*$ & 5.36 & 0.03 $\pm$ 0.02 & 69.02 \\
      &          &      &       & 0.10 $\pm$ 0.03 & 39.10 \\
$\sum$ &&&57.52&& \\ 
\hline 
\end{tabular}
\end{scriptsize}
\label{logindihd25515}
\end{center}
\end{table}
\begin{figure*}[ht]
\centering
\hbox{
\includegraphics[width=9.5cm,height=7cm]{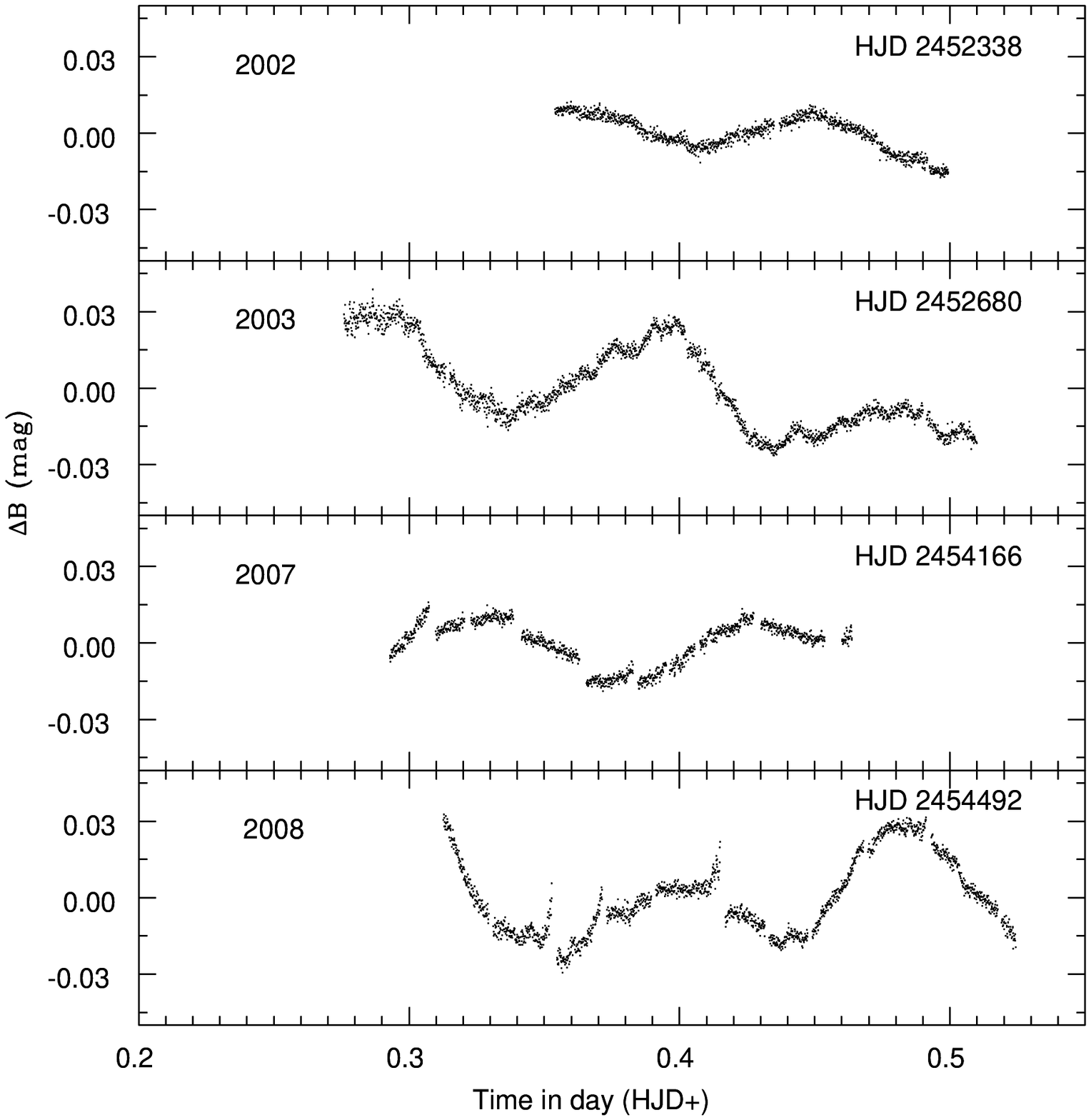}
\includegraphics[width=9.5cm,height=7cm]{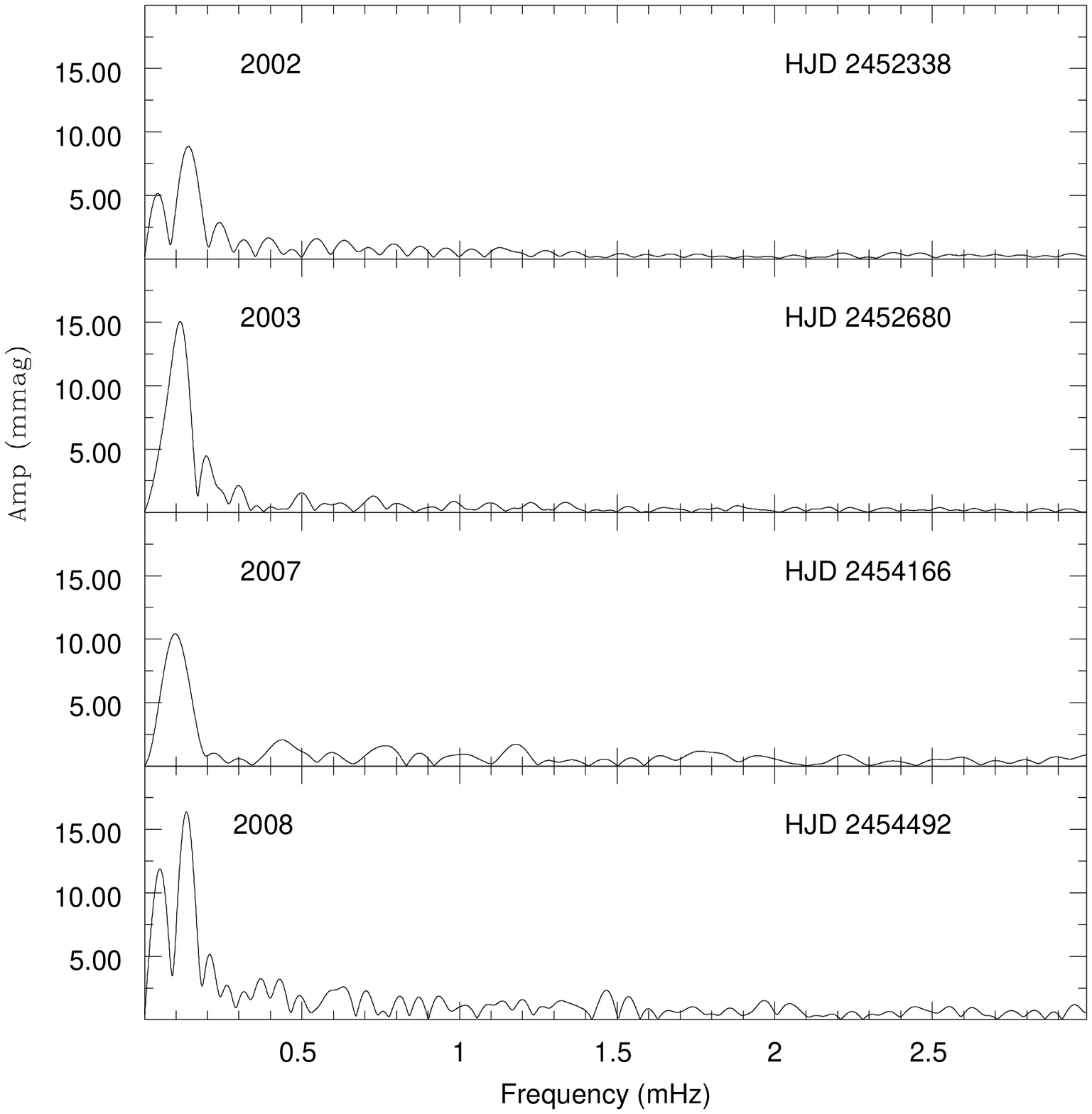}
}
\caption{ : Four typical curves (left) and corresponding amplitude spectra (right) of HD\,113878 obtained on four different observing seasons.}
\label{hd113878lc}
\end{figure*}
\begin{figure*}[ht]
\centering
\hbox{
\includegraphics[width=9.5cm,height=11cm]{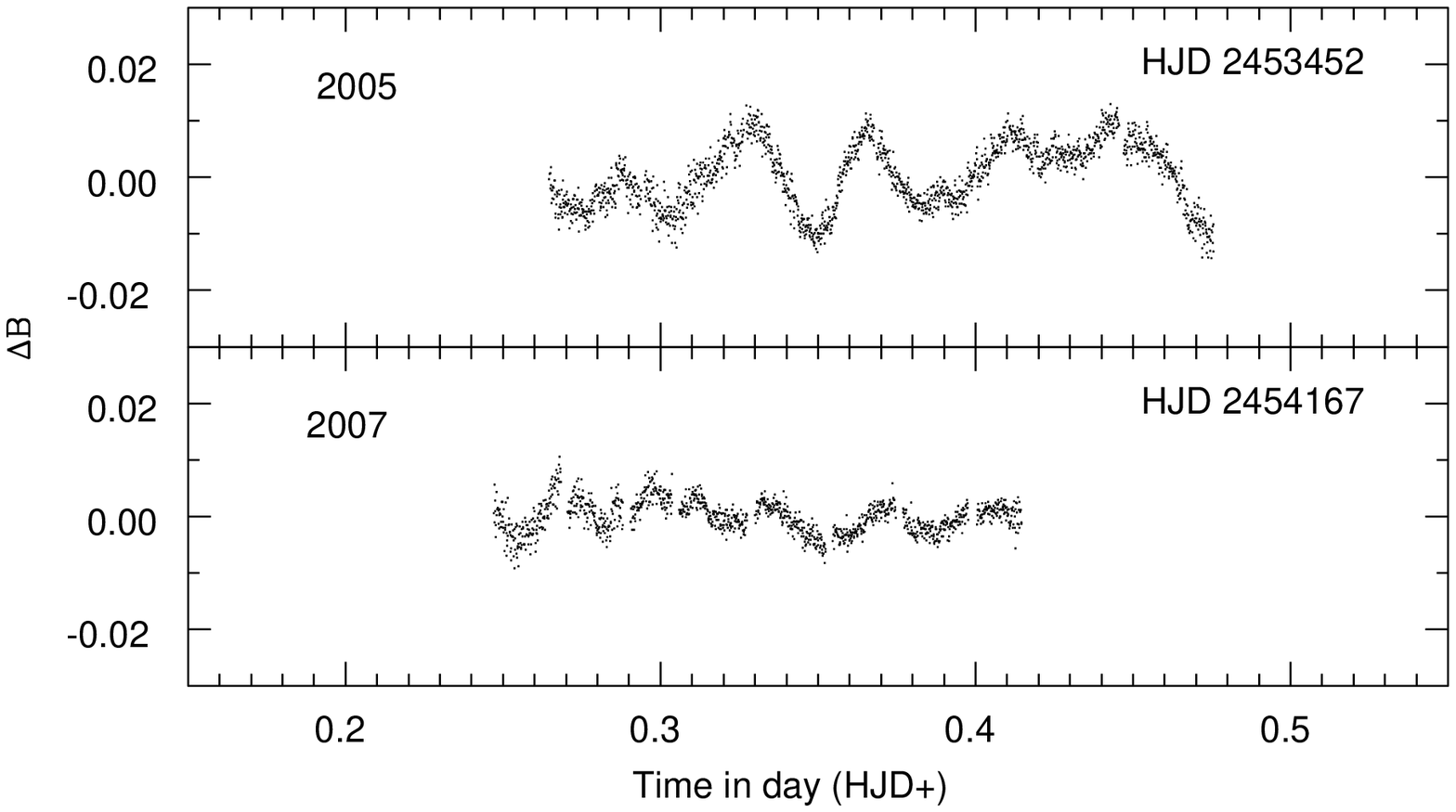}
\includegraphics[width=9.5cm,height=11cm]{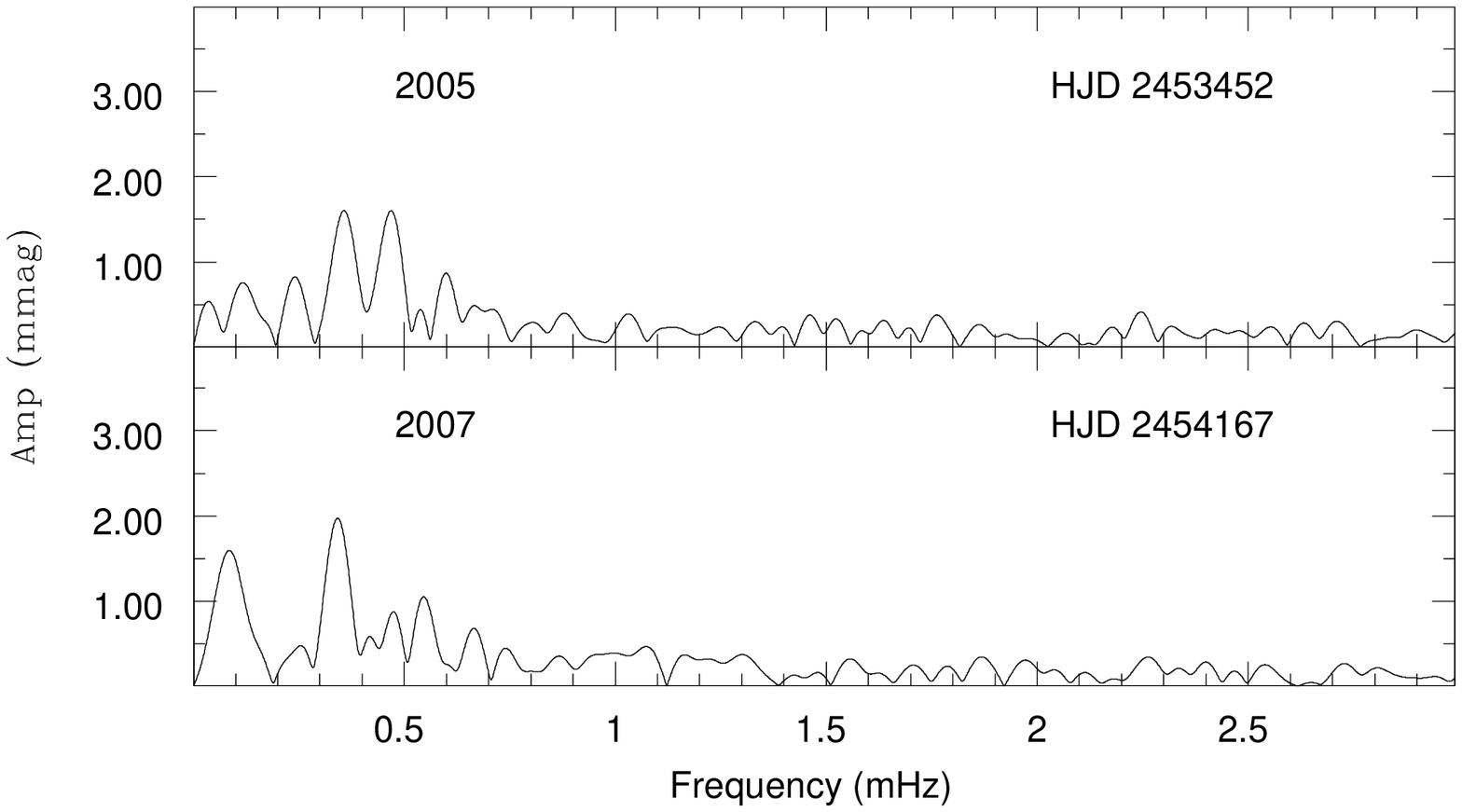}
}
\vspace*{-5.0cm}
\caption{ : Two typical curves (left) and corresponding amplitude spectra (right) of star HD\,118660 obtained on two different seasons.}
\label{hd118660lc}
\end{figure*}
Table \ref{logindihd25515} gives the complete journal of the photometric observations. 
Columns 5 and 6 of this table list the prominent frequencies and the corresponding amplitudes, respectively. A frequency analysis of the longest time-series (Dec. 20, 2007) evidences a pulsational period of 2.78-hrs ($f_1$=0.10$\pm$0.02-mHz). Another prominent frequency $f_2\approx0.03$-mHz is also visible in some amplitude spectra; however, it lies in the region where the sky transparency has a high power, further follow-up observations are therefore needed to claim the detection of this frequency. 

\begin{table}
\begin{center}
\caption{ : Journal of observations of  HD\,113878. A value $x$ in the Start Time HJD columns means Heliocentric Julian Date 2450000+$x$. The data sets marked with asterisks are used for frequency analysis to search for the additional frequencies and define the known frequencies in a better way.}
\medskip
\begin{scriptsize}
\begin{tabular}{|r|c|c|c|c|r|}
\hline
S. No. & UT & Start time & $\Delta$t  & $f_1$,$f_2$& $A_1$, $A_2$\\
   &Date & HJD & (hr) & (mHz)~~~ &mmag\\
\hline
1. & 20-01-02 & 2295.52971 & 0.66 & 0.29 $\pm$ 0.20& 5.97 \\
       &          &            &      & 0.26 $\pm$ 0.16 & 3.03 \\
2. & 14-02-02 & 2320.40533$^*$ & 1.49 & 0.13 $\pm$ 0.05 & 10.31 \\
       &          &                &   & 0.13 $\pm$ 0.04 & 2.38 \\
3. & 27-02-02 & 2333.45894 & 1.09 & 0.16 $\pm$ 0.06 & 11.90 \\
       &          &                 &   & 0.26 $\pm$ 0.06 & 3.03 \\
4. & 04-03-02 & 2338.35401$^*$ & 3.36 & 0.13 $\pm$ 0.05 & 8.87 \\
       &          &                 &  & 0.06 $\pm$ 0.04 & 3.17\\
5. & 10-01-03 & 2650.42210$^*$ & 2.22 & 0.11 $\pm$ 0.08 & 10.45 \\
       &          &                 &  & 0.24 $\pm$ 0.05 & 0.91\\
6. & 12-01-03 & 2652.35300$^*$ & 3.66 & 0.11 $\pm$ 0.04& 6.63\\
       &          &                 &  & 0.15$\pm$ 0.07 & 1.00\\
7. & 24-01-03 & 2664.32537 & 3.93 & 0.08 $\pm$ 0.05&5.10 \\
       &           &                &  & 0.16$\pm$ 0.08 & 1.94 \\
8. & 09-02-03 & 2680.27589 & 5.38 & 0.03 $\pm$ 0.02 & 15.97 \\
        &          &                &  & 0.11 $\pm$ 0.04 & 15.02 \\
9. & 06-03-07 & 4166.29280 & 3.52 & 0.09$\pm$ 0.04 & 10.42 \\
        &          &                &  & 0.15 $\pm$ 0.08 & 2.88 \\
10. & 26-01-08 & 4492.31272 & 4.68 & 0.03 $\pm$ 0.02 & 49.10\\
        &          &                &  & 0.13 $\pm$ 0.03 & 16.38 \\
11. & 18-03-08 & 4544.16985 & 5.98 & 0.04 $\pm$ 0.02 & 86.61 \\
        &          &                &  & 0.12 $\pm$ 0.10 & 26.01 \\
$\sum$ &&&35.97&& \\  
\hline
\end{tabular}
\end{scriptsize}
\label{logindihd113878}
\end{center}
\end{table}

\begin{table}
\begin{center}
\caption{: Journal of observations of HD\,118660. A value $x$ in the Start Time HJD columns means Heliocentric Julian Date 2450000+$x$.The data sets marked with asterisks are used for frequency analysis to search for the additional frequencies and define the known frequencies in a better way.}
\medskip
\begin{scriptsize}
\begin{tabular}{|r|c|c|c|c|r|}
\hline
S. No. & UT & Start time & $\Delta$t  & $f_1$,$f_2$, $f_3$& $A_1$,$A_2$,$A_3$\\
    &Date & HJD & (hr) & (mHz)~~~ &(mmag)\\
\hline
1.  & 24-02-05 & 3426.38106$^*$ & 2.98 & 0.06 $\pm$ 0.05 & 19.98 \\
        &          &               &   & 0.10 $\pm$ 0.07 & 7.02 \\
        &          &               &   & 0.28 $\pm$ 0.05 & 5.80 \\
2.        & 22-03-05 & 3452.26468$^*$ & 4.92 & 0.28 $\pm$ 0.04 & 5.08 \\
        &          &               &   & 0.11 $\pm$ 0.03 & 3.85 \\
        &          &               &   & 0.04 $\pm$ 0.04 & 2.55 \\
3. & 28-03-05 & 3458.26445 & 5.08 & 0.04 $\pm$ 0.03 & 51.21 \\
        &          &               &   & 0.09 $\pm$ 0.05 & 16.56 \\
        &          &               &   & 0.05 $\pm$ 0.03 & 15.15 \\
4. & 29-03-05 & 3459.21991 & 1.63 & 0.12 $\pm$ 0.04 & 27.85 \\
        &          &               &   & 0.23 $\pm$ 0.08 & 15.73\\
        &          &               &   & 0.48 $\pm$ 0.06 & 6.21 \\
5. & 30-03-05 & 3460.27455 & 4.54 & 0.03 $\pm$ 0.03 & 110.51 \\
        &          &               &   & 0.05 $\pm$ 0.03 & 40.86 \\
        &          &               &   & 0.10 $\pm$ 0.03 & 25.04 \\
6. & 31-03-05 & 3461.27213 & 5.03 & 0.03 $\pm$ 0.02 & 73.51  \\
        &          &               &   & 0.06 $\pm$ 0.03 & 30.92  \\
        &          &               &   & 0.10 $\pm$ 0.03 & 14.08  \\
7. & 04-04-05 & 3465.22807 & 3.86 & 0.05 $\pm$ 0.04 & 31.01 \\
        &          &               &      & 0.08 $\pm$ 0.03 & 10.85 \\
        &          &               &      & 0.13 $\pm$ 0.03 &  8.25 \\
8. & 05-01-07 & 4106.39305$^*$ & 2.59 & 0.07 $\pm$ 0.05& 6.62\\
           &            &         &     & 0.31 $\pm$ 0.06& 4.25\\
           &            &         &     & 0.13 $\pm$ 0.05 & 1.49\\
9. & 04-03-07 & 4164.25019$^*$ & 3.52 & 0.25 $\pm$ 0.08 & 3.96 \\
           &            &         &     & 0.33 $\pm$ 0.04 & 3.02 \\
           &            &         &     & 0.10 $\pm$ 0.05 & 1.24 \\
10. & 05-03-07 & 4165.33844$^*$ & 2.51 &  0.06 $\pm$ 0.05 & 6.06 \\
           &            &          &      & 0.28 $\pm$ 0.07 & 2.69 \\
           &            &           &     & 0.08 $\pm$ 0.06 & 1.49 \\
11. & 07-03-07 & 4167.24710 & 4.56 & 0.04 $\pm$ 0.03 & 19.64 \\
           &            &             &      & 0.04 $\pm$ 0.04 &4.22 \\
           &            &             &      & 0.27 $\pm$ 0.02 & 3.76 \\
$\sum$&&&41.22&& \\  
\hline 
\end{tabular}
\end{scriptsize}
\label{logindihd118660}
\end{center}
\end{table}    

\subsection{HD\,113878}
\label{hd113878}
HD\,113878 was discovered as a $\delta$-Scuti type pulsator by Joshi et al. (\cite{joshi06}). To confirm the  pulsational variability and search for additional periodicities, follow-up observations of HD\,113878 were carried out.  A total of 35.97-hrs photometric data was obtained on 11 nights between year 2002 and 2008. The left panel of Fig. \ref{hd113878lc} shows the light curves of HD\,113878 obtained on four observing seasons and the corresponding amplitude spectra are shown in the right panel. Table \ref{logindihd113878} lists the complete journal of the these observations. Columns 5 and 6 of this table list the prominent frequencies and the corresponding amplitudes. The frequency $f_1$=0.12-mHz reported by Joshi et al. (\cite{joshi06}) appears as the prominent peak in almost  all the observing runs. Therefore, these 36-hrs of photometric observations confirm the period of  pulsations in HD\,113878. Based on the available data, presently nothing can be said about other possible frequencies. 

\subsection{HD\,118660}
\label{hd118660}
The discovery of pulsations in HD\,118660 was also reported by Joshi et al. (\cite{joshi06}); it was speculated that the star might be a multi-periodic pulsating variable. To confirm this periodicity and search for additional periods, follow-up observations were carried out between year 2005 and 2007. A total of 41.22-hrs high-speed photometric data was acquired on 11 nights. Table \ref{logindihd118660} gives the journal of the observations. The left panel of  Fig. \ref{hd118660lc} shows two typical light curves obtained on two observing seasons and the corresponding amplitude spectra are shown in the right panel. The frequencies $f_1$=0.28-mHz and  $f_2$=0.11-mHz reported by Joshi et. al. (\cite{joshi06}) are indeed present in almost all the runs, so these new observations confirm the previously known frequencies.  

\subsection{Frequency Analysis of the Combined Data}
\label{combined}
\begin{figure}[ht]
\centering
\includegraphics[width=9.5cm,height=8cm]{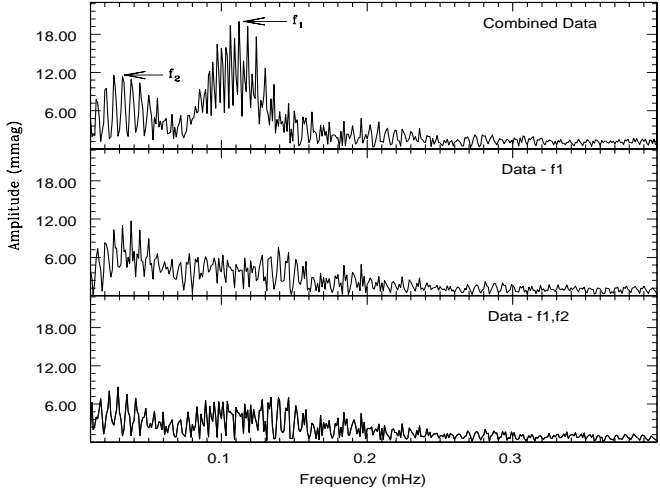}
\caption{:Amplitude spectrum of HD\,25515 after combining the time-series data marked  with asterisks in Table \ref{logindihd25515}.}
\label{hd25515combft}
\end{figure}

\begin{figure}[ht]
\centering
\includegraphics[width=9.5cm,height=8cm]{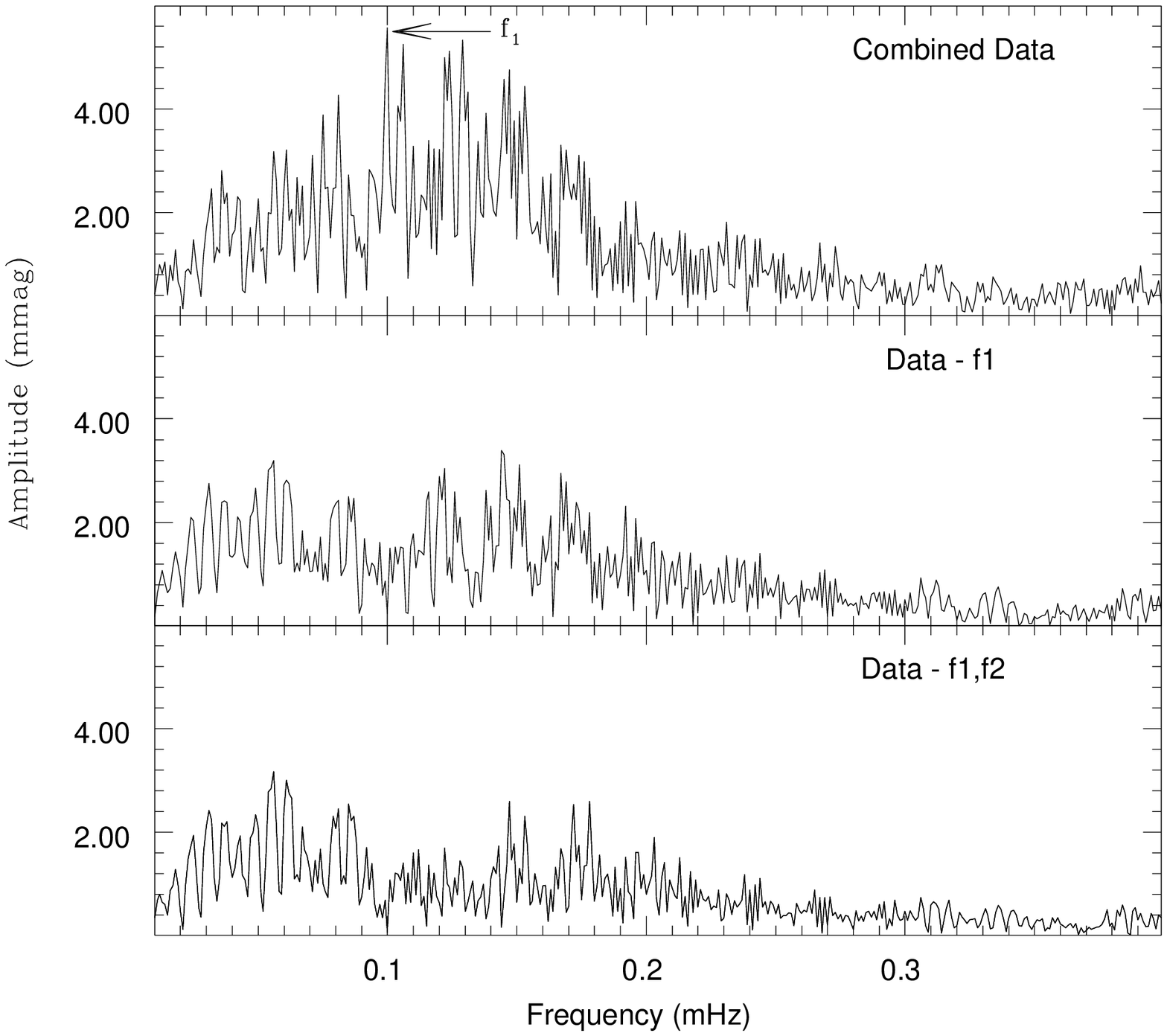}
\caption{:  Amplitude spectrum of HD\,113878 after combining the time-series data marked with asterisks in Table \ref{logindihd113878}.}
\label{hd113878combft}
\end{figure}

\begin{figure}[ht]
\centering
\includegraphics[width=9.5cm,height=9.5cm]{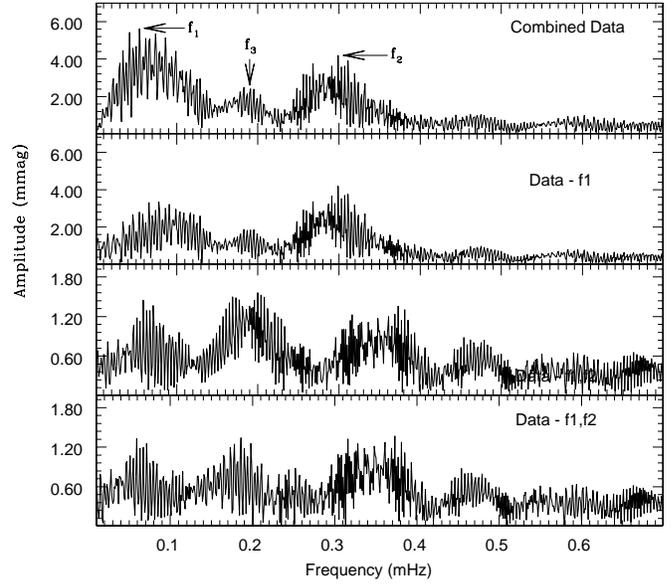}
\caption{:  Amplitude spectrum of HD\,118660 after combining the time-series data marked with asterisks in Table \ref{logindihd118660}.}
\label{h118660combft}
\end{figure}

We combined the time-series data sets of HD\,25515, HD\,113878 and HD\,118660 (Table \ref{logindihd25515}, \ref{logindihd113878} and \ref{logindihd118660}) to identify the known frequencies and to search  additional frequencies. In order to obtain a good S/N ratio in the amplitude spectra, we excluded the data sets which are contaminated by large sky transparency variation and/or have duration less than one cycle. The amplitude spectra of the combined data set of HD\,25515, HD\,113878 and HD\,118660 are shown in Fig. \ref{hd25515combft}, \ref{hd113878combft} and \ref{h118660combft}, respectively. All these Fourier transforms are ``raw'' transforms in the sense that no low-frequency filtering to remove sky transparency
fluctuations was applied to the data, other than a correction for mean extinction. The top panel of these figures are the FTs of the whole data sets mentioned in  the caption of each figure while the lower panels are the FTs after successive prewhitening by $f_1$, $f_2$, $f_3$, etc. Column 2, 3 and 4 of Table \ref{logcomb} lists the prominent frequencies,  amplitudes and phases. Two frequencies  $f_1$=0.11-mHz and $f_2$=0.03-mHz are clearly visible in the amplitude spectrum of HD\,25515. However, the frequency $f_2$ lies in the  region of sky transparency variations and hence needs confirmation. In the case of HD\,113878, the frequency $f_1$ is present in both the
combined and individual data sets and hence this frequency  is firmly confirmed. The amplitude
spectrum of HD\,118660 clearly shows the presence of three prominent
peaks at frequencies $f_1$=0.054, $f_2$=0.299 and $f_3$=0.200-mHz. The frequency  $f_1$ has a low amplitude and a low frequency so it should be taken cautiously.

The frequencies identified  using the DFT were fitted simultaneously to the combined data by linear least-squares, which assumes that the frequencies are perfectly known and adjusts the amplitudes and phases. These amplitudes and phases, along with their errors, are listed in Table \ref{logcomb}. The comparison of the amplitude and
phase determined using both  methods are consistent to each
other. We caution that all the frequencies listed in Table \ref{logcomb} are  subject to 1-day$^{-1}$ cycle count ambiguities. In order to reduce these ambiguities and eliminate them altogether a multi-site campaign is necessary. Note that another efficient way to study the pulsational variability in $\delta$-Scuti type variables is the differential CCD photometry. This technique needs nearby comparison
stars of similar magnitude and color. HD\,25515, HD\,113878 and HD\,118660 are however brighter than nine magnitude and there are no nearby comparison stars of similar magnitude and color; this renders the use of differential photometry difficult in this case. 

\begin{table}
\begin{center}
\caption{: Identification of the frequencies present in HD\,25515, HD\,113878 and HD\,118660 after combining the time-series data sets marked with asterisks in Table \ref{logindihd25515}, \ref{logindihd113878} and \ref{logindihd118660}, respectively. The phase $\phi$ corresponds to $t_0$=24553012.14302, 2452320.40523 and 2453426.38106 for HD\,25515, HD\,113878, and HD\,118660, respectively.
}
\medskip
\begin{scriptsize}
\begin{tabular}{|c|c|cc|cc|}
\hline
&&&&& \\  
Star  && \multicolumn{2}{|c|}{DFT}  & \multicolumn{2}{|c|}{LST}  \\
&&&&& \\  
\cline{1-6}
&&&&& \\  
 HD  &     {\bf $f$} & $ A $ & $\phi$ & $ A $ & $\phi$ \\ 
&&&&& \\     
\hline
&&&&& \\  
25515     & 0.112 & 20.01 & -1.482   &  22.40 $\pm$ 0.33   &  -1.513 $\pm$ 0.015  \\
                           & 0.038 & 11.68  & 2 .964 &  9.53 $\pm$ 0.33   &  -3.020 $\pm$ 0.036 \\   
&&&&& \\                        
\hline
&&&&& \\   
113878   &  0.100  & 5.57  & 0.227  &   5.60 $\pm$ 0.13   &   0.279 $\pm$ 0.023   \\
&&&&& \\    
\hline
&&&&& \\  
118660  &  0.054  & 5.62 &  2.485  &   5.90 $\pm$ 0.17   &  2.563 $\pm$ 0.028   \\
              &  0.299  & 4.18 & -1.732    &   4.23 $\pm$ 0.17   & -1.766 $\pm$ 0.039  \\
             &   0.200  & 1.56 & -0.815     &   1.77 $\pm$ 0.17   & -0.684 $\pm$ 0.094   \\
&&&&& \\  
\hline
\end{tabular}
\end{scriptsize}
\label{logcomb}
\end{center}
\end{table}

\subsection{Fundamental Parameters}
\label{fundamentalstars}

We derived some astrophysical parameters of HD\,25515, HD\,113878 and HD\,118660 using $Hipparcos$ parallaxes (van Leeuwen \cite{Lwn07}) and Simbad data. Those are listed in Table \ref{phypara02}. The calculated values for E(B-V) and the bolometric corrections (BC) for HD\,25515, HD\,113878 and HD\,118660 are 0.02, 0.20, 0.06-mag  and 0.026, 0.035, 0.035-mag, respectively. The effective temperatures of these stars were estimated  using the grids of Moon \& Dworetsky (\cite{moon85}) with typical errors of the order of 200 K .\\

The absolute magnitude $M_v$ is calculated using the relation (Cox \cite{cox99})

\begin{equation}
M_v = m_v + 5 + 5\,log\, \pi -Av,
\end{equation}

where $\pi$ is trigonometric parallax measured in arcsec,
the interstellar extinction in the V-band is $A_V$ = $R_V$E(B-V)=3.1E(B-V). The  reddening parameter E(B-V) is obtained by taking the difference of observed color (taken from Simbad data sets) and intrinsic color (estimated from Cox \cite{cox99}). 

The luminosity log $(\frac{L_\star}{L_\odot})$ were calculated using the relation

\begin{equation}
 log (\frac{L_\star}{L_\odot}) = - \frac{M_v + BC - M_{bol,\odot}} {2.5},
\end{equation}

where   $M_{bol,\odot}$ = 4.74 (Cox \cite{cox99}), and  the  bolometric correction BC is estimated using the interpolation from Flower (\cite{flower96}).\\

\begin{table}
\caption{: Astrophysical data for HD\,25515, HD\,113878 and HD\,118660 taken from Simbad Data Base , Hipparcos and using the standard relations.} 
\medskip
\begin{center}
\begin{scriptsize}
\begin{tabular}{|ccccccc|}
\hline
Star &$m_v$ & $\pi$ & d  & $M_v$ & $T_{eff}$ & log $(\frac{L_\star}{L_\odot})$ \\
HD   &(mag) & (mas) & \,\,\,(pc) & (mag) & (K) &  \\
\hline
&&&&&& \\
25515 & 8.70 &5.41$\pm$0.91 & 185$\pm$32 & 2.30$\pm$0.84 & 6823& 0.97$\pm$0.34  \\
&&&&&& \\
113878& 8.24 & 2.49$\pm$0.77 &402$\pm$137 & -0.40$\pm$1.55 & 7263& 2.04$\pm$0.62  \\
&&&&&&\\
118660 & 6.50 & 13.65$\pm$0.36 & 73$\pm$2 & 2.12$\pm$0.14 & 7500 & 1.03$\pm$0.05  \\
\hline
\end{tabular}
\end{scriptsize}
\end{center}
\label{phypara02}
\end{table}

\begin{figure}[ht]
\centering
\includegraphics[width=9.5cm,height=8.5cm]{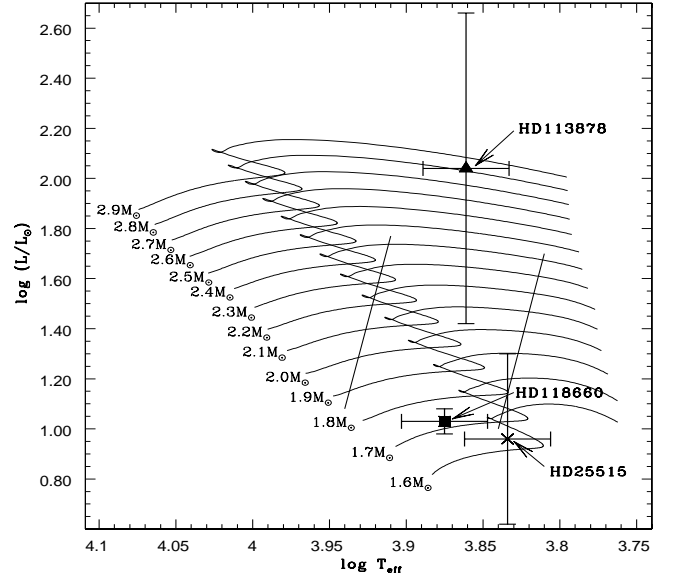}
\caption{: An HR diagram showing the positions of HD\,25515 (cross), HD\,113878 (triangle) and HD\,118660 (square) in relation to the borders of the instability strip. For the comparison evolutionary tracks of 1.6 to 2.9$M_\odot$ stars are over-plotted.}
\label{instability}
\end{figure}

The estimated positions of HD\,25515, HD\,113878 and HD\,118660 in the H-R diagram are shown in Fig. \ref{instability}. For comparison, evolutionary tracks\footnote{http://www.phys.au.dk/~jcd/emdl94/eff$\_$v6} of mass ranging from 1.6 to 2.9 $M_\odot$ are also over-plotted (Christensen-Dalsgaard \cite{christ93}).  These models are computed with the Aarhus evolution code (ASTEC), superior in terms of numerical precision and used extensively (Cunha \cite{cunha02}, Freyhammer et al. \cite{freyhammer08a},  \cite{freyhammer08b}).  For the model calculations, we used the compositions X = 0.692827 and Z = 0.02, the equation of state from Eggleton et al. (\cite{eggleton73}), opacities from OPAL92, and the convection was treated with mixing-length theory.  No convective-core overshoot, diffusion and settling  was considered to compute these models (J. Christensen-Dalsgaard, personal communication). The position of the blue (left) and red edge (right) of the instability strip are shown by two straight lines (Turcotte et al. \cite{turcotte00}).  The error in the determination of the luminosity of HD\,113878 is quite significant due to large uncertainty in the parallaxes.  From this figure it is clear that the mass of these stars are well within the range of $\delta$-Scuti stars. The H-R diagram also suggests that HD\,25515 and HD\,118660 are in the Hydrogen burning phase while HD\,113878 is in the Helium burning phase.

\section{Null Results}
\label{nullresults}
The candidate stars for which no  pulsational variability could be ascertained are addressed as null results. The null results are either constant stars, or pulsating variables with amplitude of variations  below the detection limit. Consequently, the frequency spectra of the light curves of such candidates do not show any peak to be considered as the signature of pulsations. 

It should be kept in mind that the rotational amplitude modulation and beating between frequencies may result in pulsations not being visible during  particular observations (Martinez \& Kurtz \cite{martinez94}, Handler \cite{handler04}).  So if a star is classified as a null result in a particular observation, it does not mean that the star is non-variable. For example, HD\,25515 (Sec. \ref{hd25515}) was observed for a duration of 0.78-hr on HJD2451832 and did not show any pulsational variability; it was hence classified a null result (Joshi et al. \cite{joshi06}) although it is a variable. 

 The null results are therefore important for comparison with further studies. In addition, the null result light curves are considered as essentially composed from noise, so they reflect the quality of the observation conditions. The database composed by the null results light curves is therefore very interesting for statistical studies of all noise sources (instruments, atmosphere etc.) active during the observations. 

\subsection{Fundamental Astrophysical Parameters}
\label{fundamentalnull}
Since the last survey paper (Joshi et al. \cite{joshi06}) a total of 61 candidate stars were monitored  to search for  pulsational variability  (in addition to the 3 variables described above) and few of them were observed several times. Table \ref{null} (only available in  electronic form) lists the  different astrophysical parameters of the studied sample, either available in the Simbad data base or calculated using the standard relations. For each stars, the columns of this table list respectively the HD number, right ascension $\alpha_{2000}$, declination $\delta_{2000}$, visual magnitude $m_v$, spectral type, parallax $\pi$ (van Leeuwen \cite{Lwn07}), spectral indices $b-y$,  $m_1$,  $c_1$, H$_\beta$, $\delta m_1$, $\delta c_1$, effective temperature $T_{eff}$, reddening parameter E(B-V), absolute magnitude $M_v$, luminosity log $(\frac{L_\star}{L_\odot})$ (Sec. \ref{fundamentalstars}), duration of the observations $\Delta t$, Heliocentric Julian Dates (HJD) and year of observations when the star was observed. Visual magnitude $m_v$, spectral type, parallax $\pi$, spectral indices $b-y$,  $m_1$,  $c_1$ and H$_\beta$ values are taken from the Simbad data base. $\delta m_1$ and  $\delta c_1$ are calculated using the calibration of Crawford (\cite{crawford75},\cite{crawford79}) and $T_{eff}$ is calculated using the grids of Moon \& Dworetsky (\cite{moon85}). The typical error in the estimation of the effective temperature of sample stars is about 200 K.\\





Fig. \ref{unpre01} (only available in  electronic form) shows the amplitude spectra of the light curves corrected  for extinction. As we are searching  pulsational variability of the mmag order  in the period range of a few minutes to a few hours, an amplitude  scale of 0 to 9 mmag in the y-axis and frequency scale of 0 to 5-mHz in x-axis are chosen.  The data reduction process is common for all stars so their spectra form a relatively homogeneous set of data.  Fig. \ref{pre01} (only available in  electronic form) shows the  prewhitened spectra that have been filtered for low-frequency sky transparency variations. The prewhitening strategy was to remove the high energy (above 3 mmag) low-frequency peaks (below $\approx$ 0.5-mHz).

\subsection{Analysis of the Null Results}
\label{analysisnull}
Before proceeding further, we have to correct for two informations regarding the null results presented in the previous survey paper (Joshi et al. \cite{joshi06}). Firstly, the time span of the data set in the previous paper was not 1999-2004, as mentioned in the text, but 1999-2002 (precisely 1999 November 15, to 2002 March 05). Secondly, Fig. 7 and 8 were interchanged. These figures are very similar to each other so it does not change the essence of the comments, but we nevertheless apologize to the readers and to the Editor.

\begin{figure}[t]
\centering
\includegraphics[width=9cm,height=7cm]{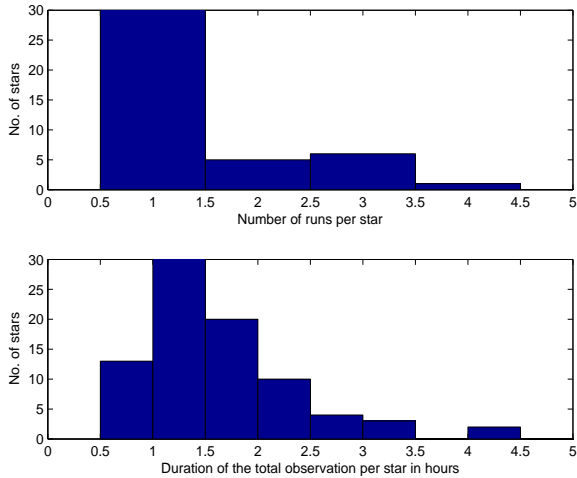}
\caption{: Statistics of the 87 runs acquired by the Nainital-Cape survey in the time range 2002-2008. Top: Distribution of the number of runs per star. Most of the stars have been observed only once. A few of them, where some variability was suspected, were observed many times. Bottom: Distribution of the total time dedicated to observe one star (in hours). Most of the stars have been observed less than 2-hrs (the mean is 1.6-hrs and the median 1.5-hrs).}
\label{stat}
\end{figure}

Fig. \ref{stat} shows the distribution of the number of runs per star, and the duration of the runs per star for the 2002-2008 data. These plots show the detection strategy of the observations : observe the candidates for a short time in order to check several targets per night.

\begin{figure}[t]
\centering
\includegraphics[width=9cm,height=7cm]{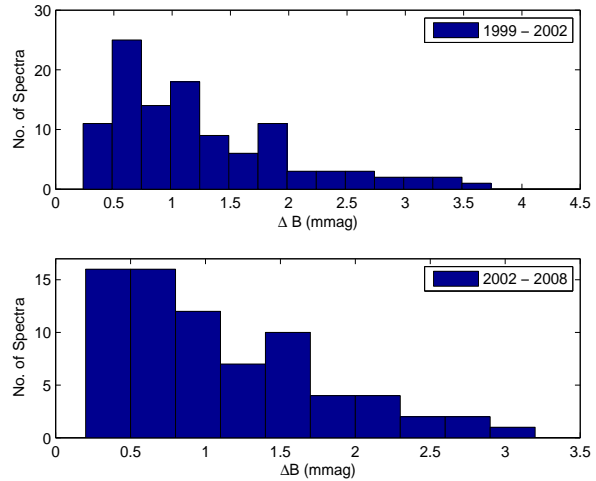}
\caption{: Distribution of the amplitude of the tallest peaks in the prewhitened spectra  for the 1999-2002 data (top) and 2002-2008 data (bottom).}
\label{deltab}
\end{figure}

Fig. \ref{deltab} shows the amplitude distribution of the tallest peak in the prewhitened spectra presented in the end of the paper. While in 2-hrs the level of the maxima of the
 noise peak created by scintillation is about 0.2-mmag for best photometric nights (Mary \cite{mary06}), it has higher values for many nights. This is mainly because  of powerful low frequency noise (sky transparency variation), that is not totally removed by prewhitening. Most light curves have their highest peak around 0.3 to 1-mmag. This range can be considered as an operational detection limit. Notice that for quite a few nights, the highest peak has a much higher amplitude. The mean of the distribution is 1.42 $\pm$ 0.14-mmag and the median 1.0-mmag. For the data set of the previous observation period (Nov. 99 to March 2002), we had for the amplitude distribution of the highest peak a mean of 1.25 $\pm$ 0.07-mmag and a median of  1.0-mmag. While the median is stable, the mean indicates a slight increase in the average power of the noise during the last 6 years with respect to the time period 1999-2002.\\

\begin{figure}[t]
\centering
\includegraphics[width=9cm,height=7cm]{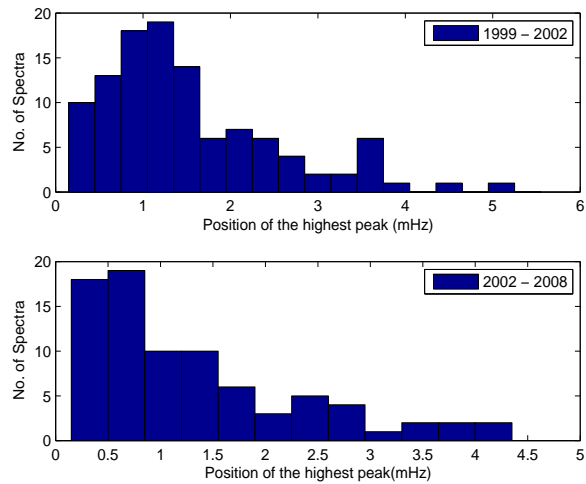}
\caption{: Distribution of the positions of the tallest peaks in the prewhitened spectra  for the 1999-2002 data (top) and 2002-2008 data (bottom).}
\label{pos}
\end{figure}

\begin{figure}
\centering
\includegraphics[width=16cm,height=20cm]{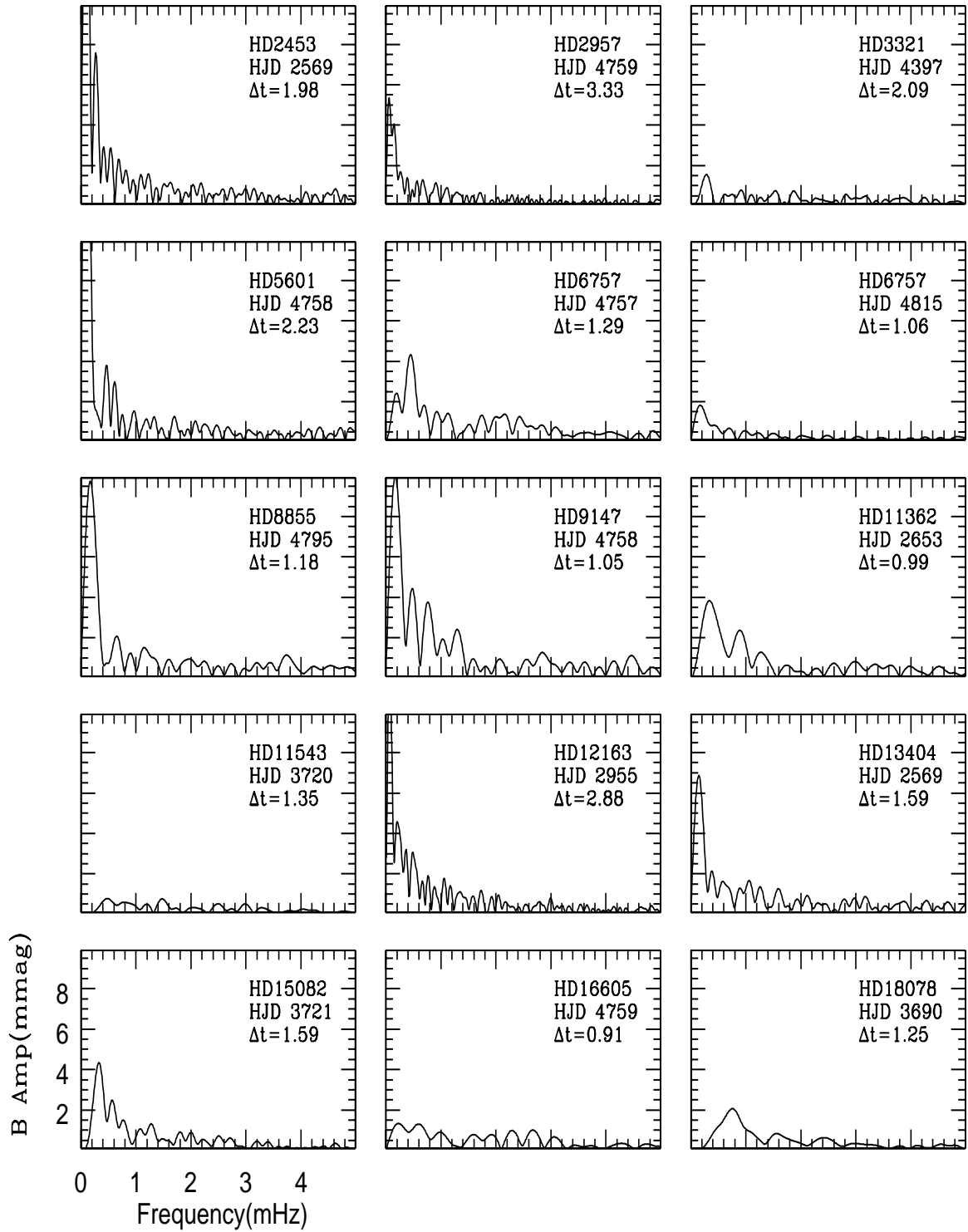}

\caption{: Null results from the Nainital-Cape survey: Examples of amplitude spectra for sample stars corrected only for extinction, and in some cases for some very long-term sky transparency variations. Each panel contains the Fourier transform of an individual light curve, covering a frequency range of $0$ to $5$\,mHz, and an amplitude range of 0 to 9 mmag. The name of the object, date of the observation in Heliocentric Julian date (HJD 245000+) and duration of the observations in hours (hr), are mentioned in each panel.}
\label{unpre01}
\end{figure}

Fig. \ref{pos}  shows the distribution in position (mHz) of the highest
peak. The distribution is not flat as would be expected from a purely scintillation noise spectrum; this is again due to sky transparency variations. The mean  and the median are respectively 1.25 $\pm$ 0.11-mHz and 0.9-mHz, against 1.52 $\pm$ 0.10-mHz and 1.3-mHz for the 1999-2002 data set. This indicates that the highest peak is shifted to the low frequency part of the spectrum, suggesting in turn that sky transparency variations (occurring at lower frequencies) are getting more powerful as before. 

Astronomers used to the Nainital site have observed  a slight decrease in the photometric  quality of Manora Peak's sky (where ARIES is located), mainly because of an increase of the human activity in the surroundings. ARIES is now in the process of installing 1.3-m and 3.6-m optical telescopes at a new astronomical site, Devasthal (longitude: $79^{o}40^{'}57{''} $ East, latitude : $29^{o}22{'}26{'}' $ North, altitude : 2420-m) by the end of year 2009 and 2012, respectively. This site has an average seeing of $\approx$ 1$^{\prime\prime}$ near ground and $\approx$ 0.65$^{\prime\prime}$ at 12-m above the ground (Sagar et al. \cite{sagar00}; Stalin et al. \cite{stalin01}). We expect these new observing facilities at Devasthal to contribute significantly to the area of asteroseismology in a very near future. 

\section{Conclusions}
\label{conclusions}
On the basis of about 58-hrs high-speed photometric data acquired on 13 nights on different observing seasons, it is found that HD\,25515 pulsates with a period of about 2.78-hrs. HD\,113878 was observed for 36-hrs on 11 nights during different seasons and no new frequency could be detected because the data are severely aliased. However, the previously estimated frequency (0.12-mHz) is present in all data sets. Similarly, in the 41-hrs time-series photometric observations of HD\,118660 taken on 11 nights, we could not identify any new frequencies other than the previously known. The fundamental astrophysical parameters of these stars were calculated using standard relations and summarized in Table \ref{phypara02}. The positions of HD\,25515, HD\,113878 and HD\,118660 in the H-R diagram indicate that all these stars are near the red-edge of the instability strip, where most  pulsating variables are found in general. The H-R diagram also shows that HD\,25515 and HD\,118660 are near  the MS, while HD\,113878 is an evolved star. We have also presented a catalogue of 61 null results stars acquired over the time scale 2002-2009, along with 87 spectra. A  statistical analysis of these null results shows, by comparison with past data, that the photometric quality of the nights at ARIES has slightly decreased during the last few years. The Nainital-Cape survey is an ongoing project, and in the light of the up-coming observational facilities, the future of asteroseismology at ARIES is quite promising.

\section{Acknowledgments}

We are thankful to the anonymous referee for many useful comments which helped us improving substantially the manuscript. The authors acknowledge Prof. R. Sagar, Prof. D. W. Kurtz, Drs. P. Martinez, S. Seetha, B. N. Ashoka, U. S. Chaubey and V. Girish for their participation in the ``Nainital-Cape-Survey''. SKT is thankful to Prof. U. S. Pandey for his critical reading. Resources provided by the electronic databases (SIMBAD, NASA's ADS and Hipparcos) are acknowledged. This work was carried out under the Indo-South Africa Science and Technology Cooperation (INT/SAFR/P-04/2002/13-02-2003 and INT/SAFR/P-3(3)2009).

\newpage
\onecolumn
\begin{scriptsize}
\begin{sidewaystable}
\caption{: Sample of stars classified as null results. The unprewhitened and prewhitened spectra of these stars are depicted in Figure \ref{unpre01} and Figure \ref{pre01}, respectively. The columns list S. No., HD number, right ascension $\alpha_{2000}$, declination $\delta_{2000}$, visual magnitude $m_v$, spectral type, parallax $\pi$, spectral indices $b-y$,  $m_1$,  $c_1$, H$_\beta$, $\delta m_1$, $\delta c_1$, effective temperature $T_{eff}$, reddening parameter E(B-V), absolute magnitude $M_v$, luminosity log $(\frac{L_\star}{L_\odot})$, duration of the observations $\Delta t$, Heliocentric Julian Dates (HJD:2450000+) and year of observation (2000+). The sign ``-'' is for those stars for which the basis data is not avilable in the archive.}
\medskip

\begin{tabular}{crccclcccccrrccccccc}
\hline\hline
S. No. & {Star}&{$\alpha_{2000}$}&{$\delta_{2000}$}&{$m_{v}$}&{Sp. }&{$\pi$}&{b-y}&{$m_{1}$}&{$c_{1}$}&{$\beta$}&{$\delta m_{1}$}&{$\delta c_{1}$}&{$T_{eff}$}&E(B-V)&{$M_{V}$}&{log $(\frac{L_\star}{L_\odot})$}&{$\Delta$t}&{HJD}&{Year}\\
& HD   & hh mm ss  & dd mm ss        & mag & Type & mas            & mag   &  mag  &  mag  &  mag &mag & mag  &   K & mag  & mag &  & hr& d&      \\\hline
1. &  2453   &00 28 28&+32 26 16 &6.91& A2p    & 4.51 $\pm$0.47& 0.022 & 0.244 & 0.887 & 2.854&-0.038& 2.016&  9125&0.03 &0.088&1.896 &1.98 & 2569&02 \\
2. & 2957   &00 32 44&-13 29 13 &8.50& Ap     & 3.88 $\pm$0.84&-0.008 & 0.209 & 0.991 & 2.880&-0.009& 0.061&  8350&0.03 &1.351&1.351&3.33 & 4759&08 \\
3. & 3321   &00 36 22& +33 38 39&8.40& A3     & 6.34 $\pm$0.90&      -&      -&      -&     -&     -&     -&          -& 0.22 &1.728&  - &2.09 & 4397&07 \\
4. & 5601   &00 57 32& -10 28 33&7.63& A0     & 3.58 $\pm$0.47&-0.056 &0.219  & 0.782&      -&     -&     -&          -& 0.00 &0.461&  - &2.23 & 4758&08 \\
5. & 6757   &01 08 53& +45 12 27&7.70& A0Vp   & 7.71 $\pm$1.25&      -&      -&      -&     -&     -&     -&          -& 0.42 &0.833&  - &1.29 & 4757&08 \\
&     &        & 	   &    &        &      	 &       &       &       &      &      &      &        &      &             &     &1.06 & 4815&08 \\
6. & 8855   &01 28 06& +43 52 39&8.20&A1p     & 1.84 $\pm$0.68&      -&      -&      -&     -&     -&          -&     -&0.00  &-0.166&  -  &1.18 & 4795&08 \\
7. & 9147   &01 30 56& +45 21 56&9.38&A0p     &              -&    	-&      -&      -&     -&     -&     -&        -&0.11 & -&   - &1.05 & 4758&08 \\
8. & 11362  &01 52 15& +26 10 05&8.63&A5      &              -& 0.242 &  0.227&  0.754& 2.768&-0.036& 0.038&   7411&0.19  & -&  -  &0.99 & 2653&03 \\
9. & 11543  &01 55 10& +59 30 30&8.14& Am     &  9.36$\pm$1.03& 0.144 &  0.214&  0.857& 2.846&-0.007&-0.005&   8091&0.08  &2.748&0.787&1.35 & 3720&05 \\
10. & 12163  &02 00 16& +37 38 01&8.24& F0III  &              -& 0.204 &  0.198&  0.666& 2.748&-0.012& 0.036&   7261&0.04&   -&   - &2.88 & 2955&03 \\
11. & 13404  &02 11 54& +36 57 58&8.75& A2     &  6.33$\pm$1.07& 0.212 &  0.262&  0.600& 2.809&-0.058&-0.198&  7913&0.30&1.827&1.154&1.59 & 2569&02 \\
12. & 15082  &02 26 51& +37 33 02&8.30& A5     &  8.65$\pm$0.80& 0.166 &  0.204&  0.720& 2.768&-0.013& 0.004&   7418&0.00&3.171&0.614&1.59 & 3721&05 \\
13. & 16605  &02 40 59& +42 52 16&9.63& F7V    &              -& -0.021&  0.222&  0.774& 2.816&-0.014& 0.024&   7883&0.00&   -&  -  &0.91 & 4759&08 \\
14. & 18078  &02 56 32& +56 10 41&8.30& A0p    &  0.64$\pm$0.73&  0.087&  0.251&  1.079& 2.831&-0.044& 0.243&   7947&0.22 &-3.351&3.225&1.25 & 3690&05 \\
       &        &          &    &        &               &       &       &       &      & &      &      &            &   &  &    &1.73 & 3719&05 \\ 
       &        &          &    &        &               &       &       &       &      & &      &      &            &   &  &    &0.76 & 4815&08 \\
15. & 19653  &03 12 46& +60 48 03&8.87& B9p    &  2.67$\pm$1.08& 0.170 & 0.131 &  1.045&     -&     -&     -&          -&0.29 &0.073&  -  &0.95 & 4815&08 \\    
16.  & 19712  &03 10 18& -01 41 41&7.35& B9     &  6.91$\pm$0.59& -0.060& 0.227 &  0.860& 2.858&-0.022&-0.026&     11005&0.05 &1.392 &1.512&1.58 & 4757&08 \\
17. & 21476  &03 32 01& +67 35 08&7.64& F0     &  8.74$\pm$1.16& 0.231 & 0.155 &  0.591& 2.707& 0.018& 0.039&  6929&0.01 &2.304&0.963&1.09 & 3659&05 \\
18. & 25154  &03 59 48& -00 01 13&9.88& A5     &  6.74$\pm$1.41& 0.429 & 0.236 &  0.693& 2.783&-0.039&-0.053&   7591&0.39&2.814&0.757&1.23 & 4397&07 \\
19. & 25499  &04 05 39& +53 04 28&8.05& F0     &              -& 0.225 & 0.238 &  0.690& 2.744&-0.053& 0.067&   7217&0.08&   -&   -&1.21 & 2569&02 \\
       &        &          &    &        &               &       &       &       &      & &      &      &            &   &  &    &2.13 & 2954&03 \\
       &        &          &    &        &               &       &       &       &      & &      &      &            &   &  &    &1.61 & 2954&03 \\
20. & 27404  &04 20 37& +28 53 31&7.95&A0      &  4.40$\pm$1.15& 0.183 & 0.114 &  0.788& 2.787& 0.085& 0.034&  7570&0.24&0.423&1.713&1.01 & 4758&08 \\
21. & 27716  &04 24 00& +35 12 57&7.88& F0     &  8.87$\pm$0.67& 0.228 & 0.216 &  0.787& 2.777&-0.021& 0.106&   7481&0.09&2.341&0.946&1.76 & 3400&04 \\
22. & 32428  &05 04 37& +32 19 13&6.62& A4m    & 10.28$\pm$1.06& 0.171 & 0.210 &  0.819& 2.770&-0.018& 0.099&   7406&0.00&1.540&1.210&1.26 & 3720&05 \\ 
23. & 32642  &05 05 32& +19 48 24&6.44& A5m    &  7.26$\pm$0.81& 0.124 & 0.211 &  1.014& 2.867&-0.008& 0.110&   8300&0.09&0.466&1.704&2.16 & 3660&05 \\
       &        &          &    &        &               &       &       &       &      & &      &      &            &   &  &    &1.36 & 3721&05 \\
       &        &          &    &        &               &       &       &       &      & &      &      &            &   &  &    &1.74 & 3748&06 \\
       &        &          &    &        &               &       &       &       &      & &      &      &            &   &  &    &1.70 & 3749&06 \\
24. & 35450  &05 28 24& +58 40 30&8.18& A3     &  7.42$\pm$0.87&      -&      -&      -&     -&     -&     -&          -&0.21&1.881&   -&1.04 & 4396&07 \\
       &        &          &    &        &               &       &       &       &       & &      &      &           &  &   &    &1.49 & 4397&07 \\
25. & 38817  &05 50 37& +44 00 41&7.56& A2     &  7.27$\pm$0.76&0.066  & 0.217 & 0.942 & 2.860&-0.012& 0.052&   8209&0.11&1.527&1.277&1.39 & 2563&02 \\
       &        &          &    &        &               &       &       &       &      & &      &      &            &   &  &    &0.73 & 2563&02 \\
       &        &          &    &        &               &       &       &       &      & &      &      &            &   &  &    &3.18 & 2952&03 \\
       &        &          &    &        &               &       &       &       &      & &      &      &            &   &  &    &4.25 & 2953&03 \\
       &        &          &    &        &               &       &       &       &      & &      &      &            &   &  &    &4.28 & 2976&03 \\
       &        &          &    &        &               &       &       &       &      & &      &      &            &   &  &    &1.63 & 4795&08 \\
       &        &          &    &        &               &       &       &       &      & &      &      &            &   &  &    &0.78 & 4815&08 \\
26. & 38823  &05 48 25& -00 45 34&7.32&A5      & 10.26$\pm$0.84& 0.151 & 0.301 & 0.580 &     -&     -&     -&          -&0.22   & 1.694& -&1.32 & 4758&08 \\
27. & 40759  &06 00 45& -03 53 44&8.56&A0      &  4.63$\pm$0.92& -0.013& 0.226 & 0.899 &     -&     -&     -&         -&0.05&1.733&   -&1.12 & 4795&08 \\
28. & 43058  &06 16 21& +43 39 56&9.15& A3     &              -&      -&      -&      -&     -&     -&     -&          -&0.15&  -&   -&1.82 & 3720&05 \\\hline
\end{tabular}
\label{null}
\end{sidewaystable}
\end{scriptsize}

\newpage
\addtocounter{table}{-1}
\begin{scriptsize}
\begin{sidewaystable}
\caption{:~Continued}
\medskip
\begin{tabular}{crccclcccccrrccccccc}
\hline\hline
S. No. & {Star}&{$\alpha_{2000}$}&{$\delta_{2000}$}&{$m_{v}$}&{Sp. }&{$\pi$}&{b-y}&{$m_{1}$}&{$c_{1}$}&{$\beta$}&{$\delta m_{1}$}&{$\delta c_{1}$}&{$T_{eff}$}&E(B-V)&{$M_{V}$}&{log $(\frac{L_\star}{L_\odot})$}&{$\Delta$t}&{HJD}&{Year} \\
& HD   &  hh mm ss    & dd mm ss               & mag     & Type & mas    & mag &  mag     &  mag    &   mag   &  mag           & mag            & K          & mag  & mag&   & hr        &  d    &  \\\hline
29. & 43508  &06 20 04& +56 55 31&8.86& F0     &  4.32$\pm$1.29&  0.206&  0.215&  0.844& 2.790&-0.016& 0.140& 7600&0.01 &2.006&1.080&1.34 & 3690&05 \\
       &        &          &    &        &               &       &       &       &      &      & &      &          &    & &    &1.31 & 3721&05 \\
30. & 44738  &06 23 54& +14 06 49&7.90& A2     &  2.68$\pm$0.93& -0.071&  0.244&  0.807& 2.799&-0.041& 0.029& 7650&0.01& 0.009&1.879&2.97 & 3718&05 \\
       &        &          &    &        &      	 &       &       &       &      &      &      & &          &    & &    &2.20 & 3719&05 \\
31. & 45784  &06 30 45& +29 49 42&8.11& F2     &  6.28$\pm$1.01& 0.206 &  0.234&  0.745& 2.771&-0.041& 0.074& 7438&0.00&2.255&0.980&2.98 & 2651&03 \\
32. & 49713  &06 49 44& -01 20 23&7.32& B9p    &  5.01$\pm$0.70& -0.051&  0.184&  0.655& 2.747&     -&     -&     12719&0.26&0.013&2.064&1.61 & 4795&08 \\
33. & 52069  &07 01 22& +34 59 52&8.79& A5     &              -&  0.199&  0.220&  0.712& 2.757&-0.033& 0.018& 7350&0.18 &  -&   -&1.19 & 3397&04 \\
34. & 57558  &07 23 42& +42 18 56&9.07& A5     &              -&  0.176&  0.229&  0.730& 2.779&-0.033&-0.008& 7600&0.15 &  -&   -&2.37 & 3459&04 \\
       &        &          &    & 	 &               &       &       &       &      &      &      & &          &   &  &    &2.03 & 3464&04 \\  
       &        &          &    &        &               &       &       &       &      &      & &      &          &   &  &    &2.67 & 3465&04 \\
35. & 59622  &07 31 46& +23 38 53&8.69& F0     &      	-&  0.171&  0.246&  0.781& 2.780&-0.050& 0.095& 7507&0.00 &  -&   -&1.39 &2654&03 \\
36. & 64534  &07 55 48& +35 58 22&9.18& F0     &  2.31$\pm$1.18&  0.098&  0.283&  0.871& 2.862&-0.061& 0.039& 8228&0.00&1.308&1.366&1.06 &2662&03 \\
37. & 64561  &07 54 34& +01 34 13&8.24& A3     &  8.02$\pm$0.93&  0.169&  0.278&  0.732& 2.783&-0.081&-0.014& 7548&0.25 &1.986&1.088&1.26 &2649&03 \\
38. & 66350  &08 03 01& -02 43 49&8.68& A0     &  1.73$\pm$0.92& -0.030&  0.197&  1.050&     -&     -&     -&        -&0.00 &-0.129&   -&0.85 &4815&08 \\
39. & 68703  &08 14 11& +17 40 33&6.47& A0     & 10.53$\pm$0.51&  0.174&  0.221&  0.813& 2.775&-0.027& 0.083& 7200&0.31&0.618&1.635&0.75 &3687&05 \\
       &        & 	   &    &        &      	 &       &       &       &      &      &      & &          &   &  &    &1.47 &3688&05 \\
       &        & 	   &    &        &      	 &       &       &       &      &      &      & &          &   &  &    &1.08 &3718&05 \\
40. & 71973  &08 36 49& +74 43 25&6.31& A2m    & 13.21$\pm$0.50&  0.169&  0.236&  0.766& 2.767&-0.045& 0.052& 7400&0.23& 1.202&1.401&1.57 &3721&05 \\
41. & 72943  &08 36 08& +15 18 49&6.32& F0IV   & 12.86$\pm$0.45&  0.211&  0.186&  0.732& 2.720&-0.009& 0.152& 7000&0.00 &2.207&1.001&1.43 &3812&06 \\
42. & 73093  &08 38 17& +43 10 10&9.61& A5     &              -&  0.080&  0.248&  0.963& 2.862&-0.043& 0.069& 8300&0.00 &   -&   -&1.42 &3422&04 \\    
43. & 73618  &08 39 56& +19 33 11&7.30& Am     &  3.62$\pm$1.05&  0.105&  0.230&  0.967& 2.846&-0.023& 0.105& 8100&0.04 &-0.030&1.899&1.51 &3720&05 \\    
44. & 73619  &08 39 58& +19 32 29&7.52& Am     &  3.62$\pm$1.05&  0.143&  0.237&  0.829& 2.823&-0.031& 0.004& 7900&0.09 &0.035&1.871&0.64 &3689&05 \\    
45. & 73709  &08 40 21& +19 41 12&7.68& F2III  &  4.58$\pm$0.57&  0.099&  0.258&  0.938& 2.843&-0.042& 0.140& 8100&0.00 &1.418&1.319&1.38 &3719&05 \\    
       &        &          &    &        &               &       &       &       &      &      & &      &          &   &  &    &1.10 &3748&06 \\
46. & 77314  &09 01 58& +02 40 16&7.14& Ap     &  6.78$\pm$0.91&      -&      -&      -&     -&     -&     -&        -&0.01&1.265&   -&3.12 &4549&08 \\
       &        & 	   &    &        &      	 &       &       &       &      &      &      & &          &   &  &    &1.49 &4550&08 \\
       &        & 	   &    &        &      	 &       &       &       &      &      &      & &          &   &  &    &1.48 &4551&08 \\
47. & 86170  &09 56 45& -02 17 20&8.42&A2      &  3.20$\pm$0.98&  0.074&  0.226&  0.929&     -&     -&     -&        -&0.12&0.574&   -&0.91 &4815&08 \\
48. & 86458  &10 01 32& +68 43 05&8.05& F0     &  6.28$\pm$0.95&  0.248&  0.220&  0.673& 2.711&-0.046& 0.111& 6900&0.08 &1.792&1.168&0.84 &3397&04 \\
49. & 94401  &10 53 52& +02 39 28&8.09& F0     &  6.99$\pm$0.71&  0.581&  0.382&  0.441& 2.711&-0.208&-0.121&7139&0.00 &2.312&0.958&1.42 &3458&04 \\
50. & 95256  &11 01 06& +63 25 16&6.37& A2m    & 11.19$\pm$0.66&  0.083&  0.240&  0.965& 2.845&-0.033& 0.105& 8200&0.12 &1.614&1.242&2.08 &3719&05 \\
       &        & 	   &    &        &               &       &       &       &      &      &      &   &       &    &  &  &1.99 &3721&05 \\
       &        & 	   &    &        &               &       &       &       &      &      &      & &          &   &  &    &0.98 &3812&06 \\
51. & 96003  &11 04 33& +12 40 01&6.87& A3p    &  6.06$\pm$0.49& -0.020&  0.228&  0.993& 2.893&-0.030& 0.037& 9795&0.04& 0.655&1.636&0.55 &3717&05 \\
52. & 96528  &11 07 40& +23 19 25&6.49& A5m    & 12.91$\pm$0.41&  0.093&  0.210&  0.914& 2.870&-0.008& 0.004& 8300&0.01&2.001&1.090&1.88 &3718&05 \\
53. & 110248 &12 40 35& +30 22 39&7.65& Am     &  6.59$\pm$0.67&  0.181&  0.251&  0.789& 2.808&-0.047&-0.007& 7800&0.15 &1.279&1.372&1.09 &3397&04 \\
54. & 112097 &12 53 50& +12 25 06&6.25& A7III  & 16.42$\pm$0.80&  0.175&  0.176&  0.747& 2.750& 0.009& 0.067&7300&0.08&2.245&0.984&2.37 &3812&06 \\
55. & 132739 &15 00 19& +13 19 17&8.59& F0p    &  5.49$\pm$1.02&  0.226&  0.169&  0.703& 2.719& 0.008& 0.125& 7000&0.03&2.195&1.006&1.30 &3838&06 \\ 
56. & 136403 &15 19 30& +32 30 54&6.33& A2m    & 11.50$\pm$0.35&  0.135&  0.209&  0.845& 2.817&-0.004& 0.031& 7800&0.17&1.122&1.435&2.08 &3182&06 \\
57. & 141675 &15 47 38& +55 22 36&5.88& A3m    & 13.55$\pm$0.32&  0.136&  0.248&  0.866& 2.823&-0.042& 0.041& 7900&0.15&1.078&1.453&1.58 &3481&04 \\
58. & 158116 &17 26 14& +29 27 20&7.68& Am     &  4.32$\pm$0.81&  0.156&  0.255&  0.908& 2.792&-0.054& 0.144& 7600&0.14&0.423&1.713&1.70 &3481&04 \\ 
59. & 190145 &19 58 59& +67 28 20&7.56& A2p    &  6.02$\pm$0.46&  0.129&  0.260&  0.838& 2.841&-0.052&-0.014& 8100&0.21&0.807&1.564&1.00 &3518&04 \\
60. & 198263 &20 47 24& +48 51 08&8.16& F0     &              -&  0.217&  0.267&  0.684& 2.764&-0.076& 0.026& 7400&0.05&  -&   -&1.83 &3541&04 \\
61. & 213143 &22 29 06& +21 23 35&7.75& Fm     &  4.00$\pm$0.73&  0.229&  0.246&  0.729& 2.753&-0.058& 0.090& 7300&0.07&0.543&1.665&1.66 &3690&05 \\\hline 
\end{tabular}
\end{sidewaystable}
\end{scriptsize}

\begin{center}
\vspace*{8.1cm}
{\bf \huge Online Material}
\end{center}
\newpage

\addtocounter{figure}{-1}
\begin{figure}
\centering
\includegraphics[width=16cm,height=19cm]{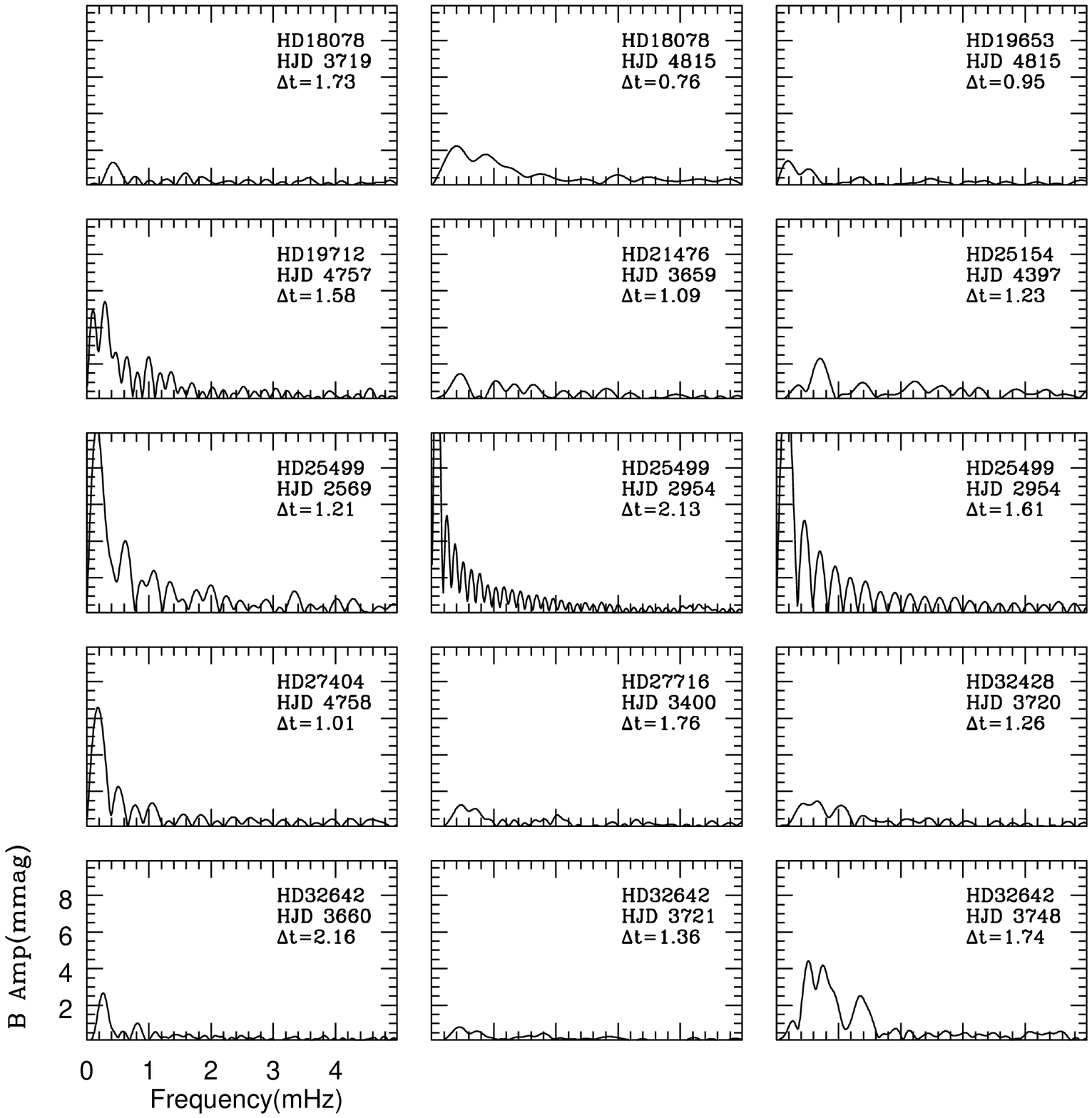}
\caption{: Cont.}
\end{figure}

\addtocounter{figure}{-1}
\begin{figure}
\centering
\includegraphics[width=16cm,height=19cm]{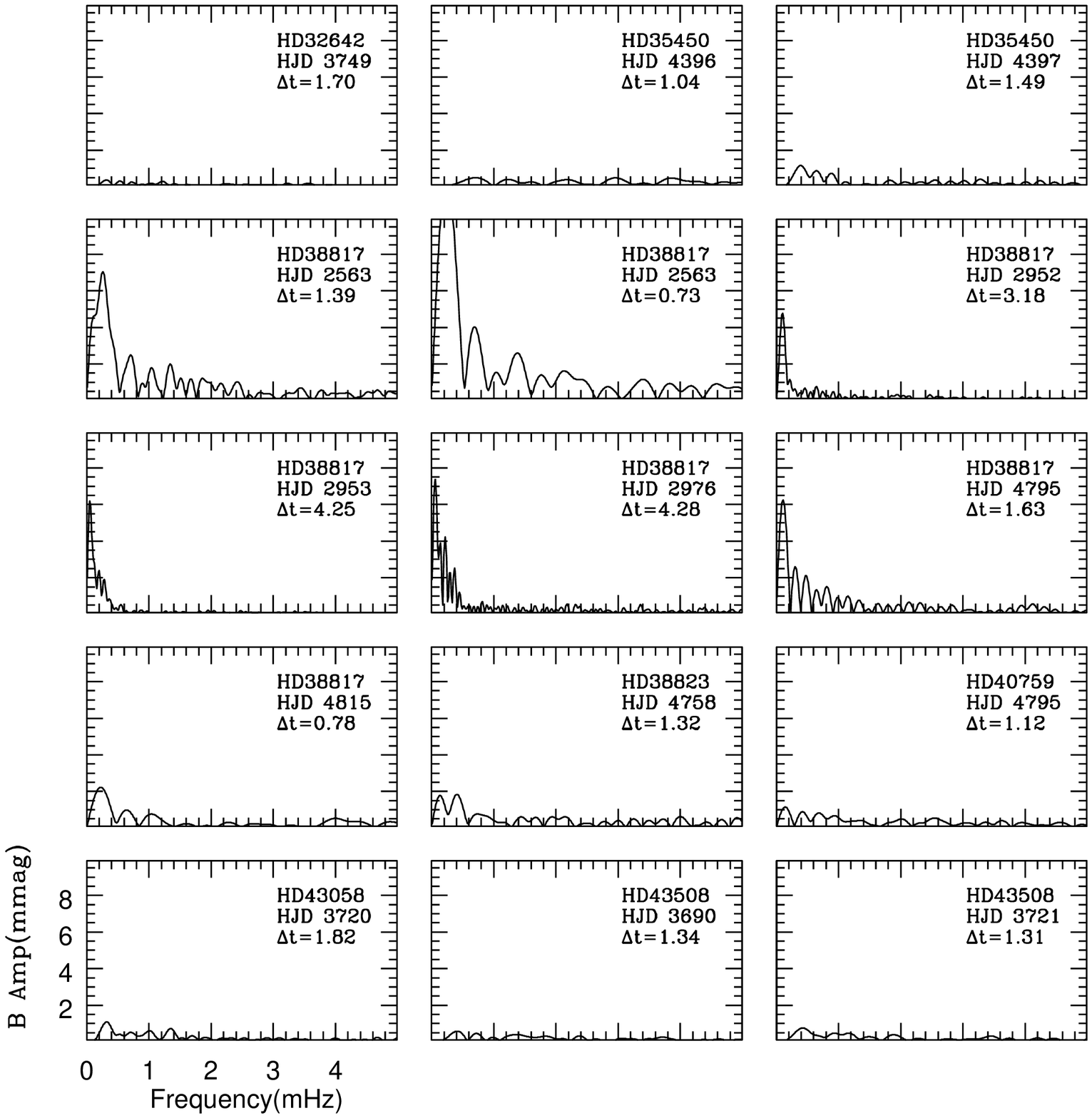}
\caption{: Cont.}
\end{figure}

\addtocounter{figure}{-1}
\begin{figure}[ht]
\centering
\includegraphics[width=16cm,height=19cm]{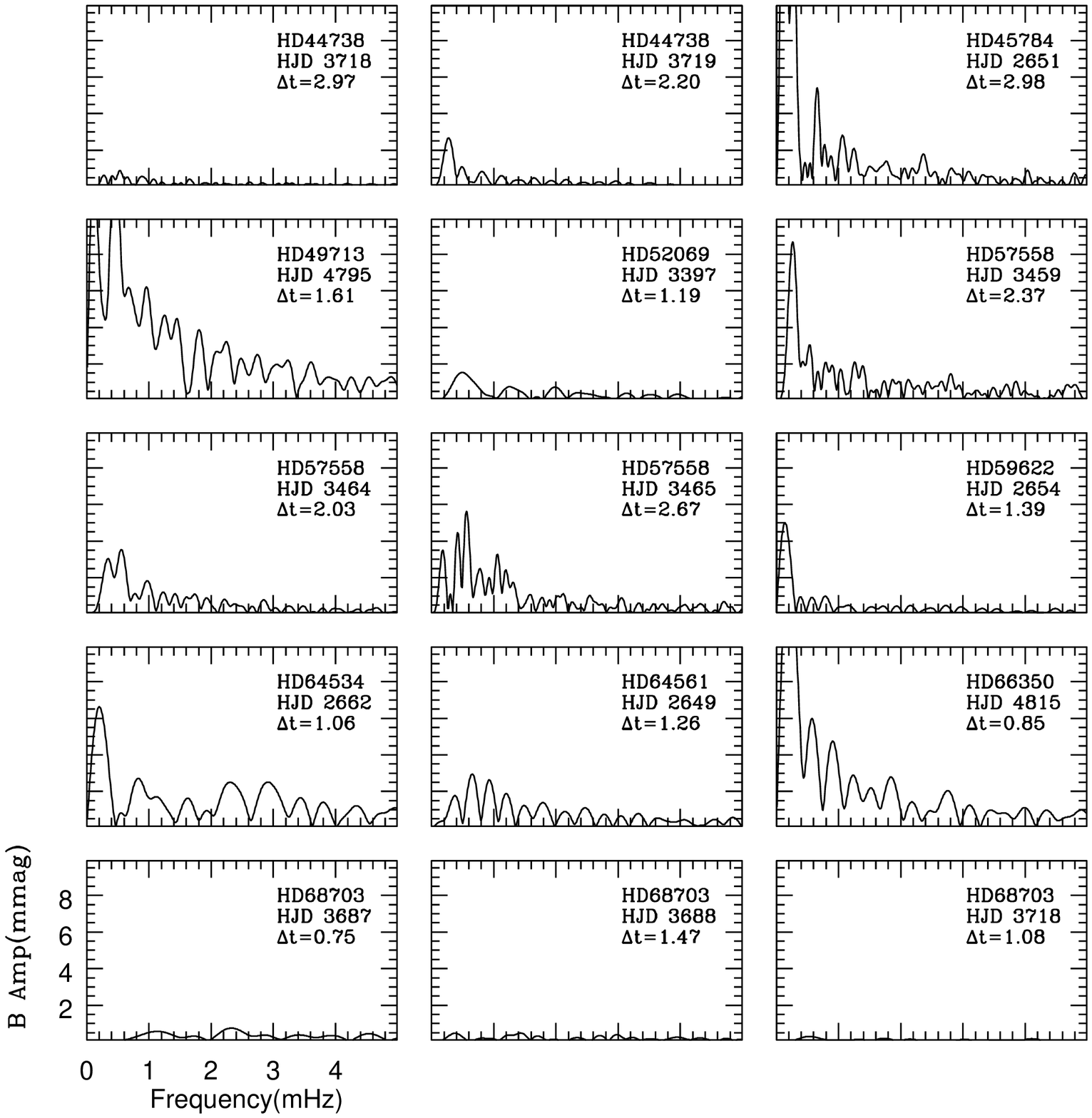}
\caption{: Cont.}
\end{figure}

\addtocounter{figure}{-1}
\begin{figure}
\centering
\includegraphics[width=16cm,height=19cm]{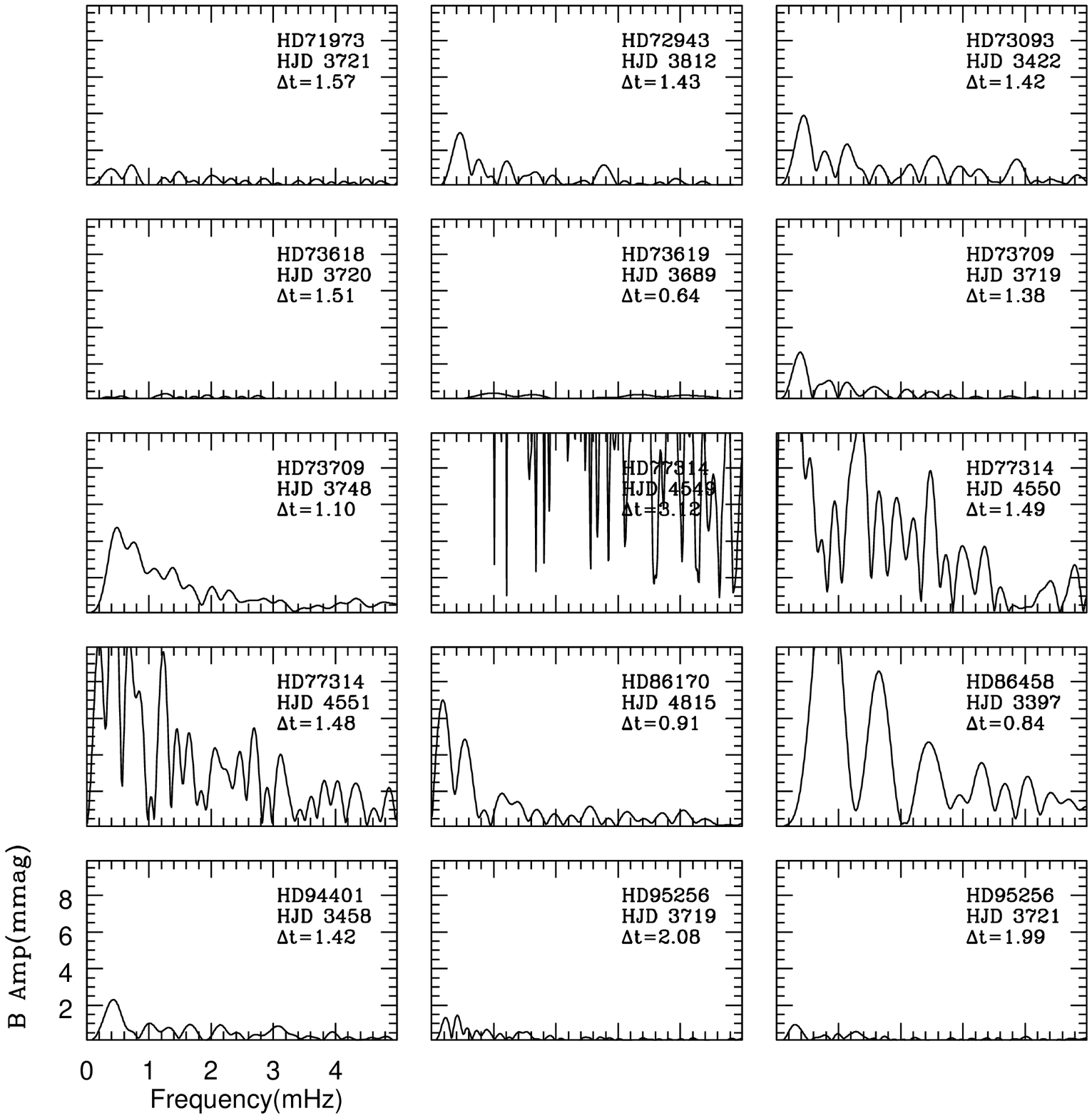}
\caption{: Cont.}
\end{figure}

\addtocounter{figure}{-1}
\begin{figure}
\centering
\includegraphics[width=16cm,height=19cm]{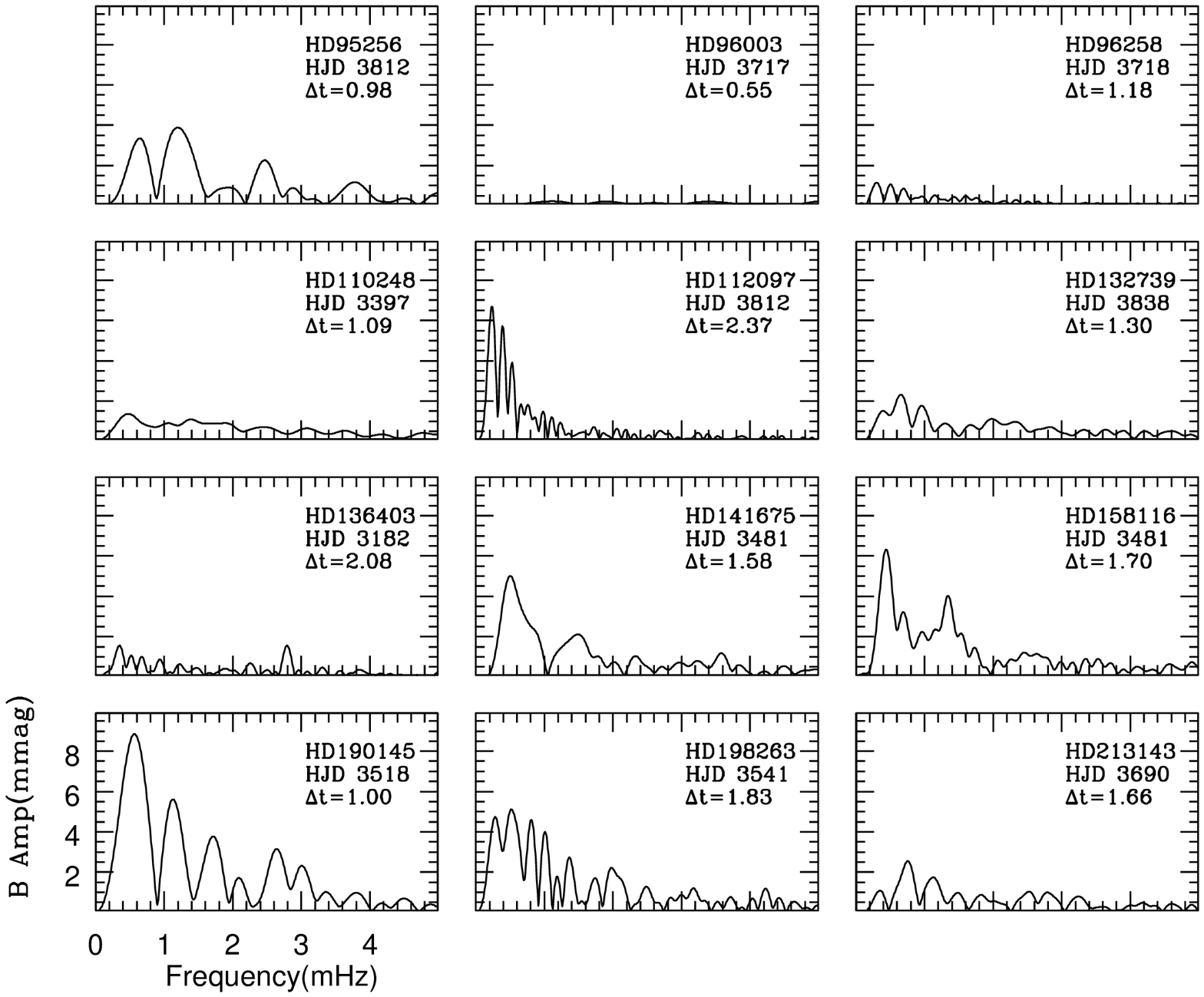}
\caption{: Cont.}
\end{figure}

\begin{figure}
\centering
\includegraphics[width=16cm,height=19cm]{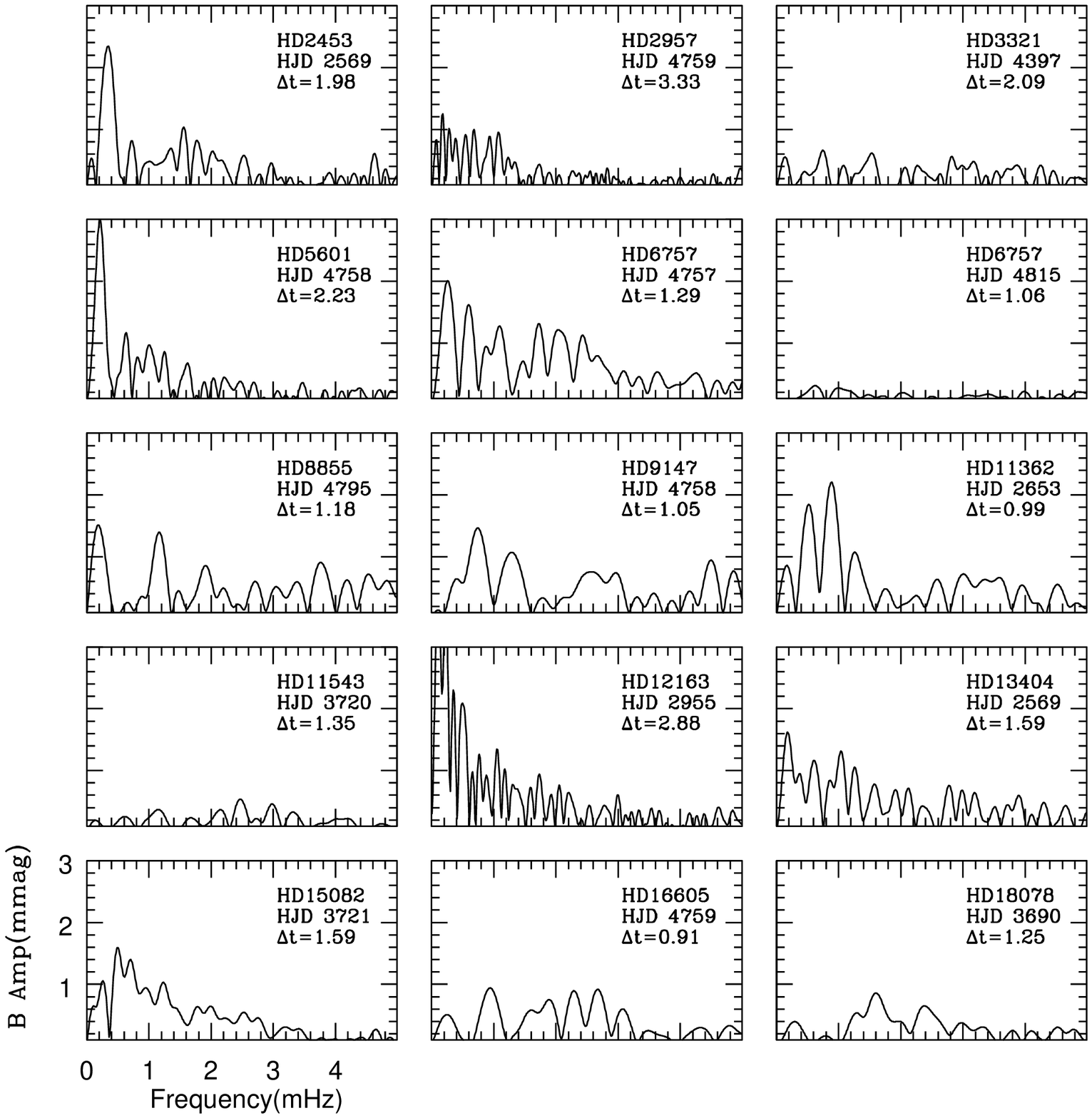}
\caption{: Null results from the Nainital-Cape survey: Examples of prewhitened
amplitude spectra for sample stars. Each
panel contains the Fourier transform of an individual light curve, covering a frequency
range of $0$ to $5$\,mHz, and an amplitude range of 0 to 3 mmag. The name of the object,
date of the observation in Heliocentric Julian date (HJD 245000+) and the length duration in hours (hr), are mentioned in each panel.}
\label{pre01}
\end{figure}

\addtocounter{figure}{-1}
\begin{figure}
\centering
\includegraphics[width=16cm,height=19cm]{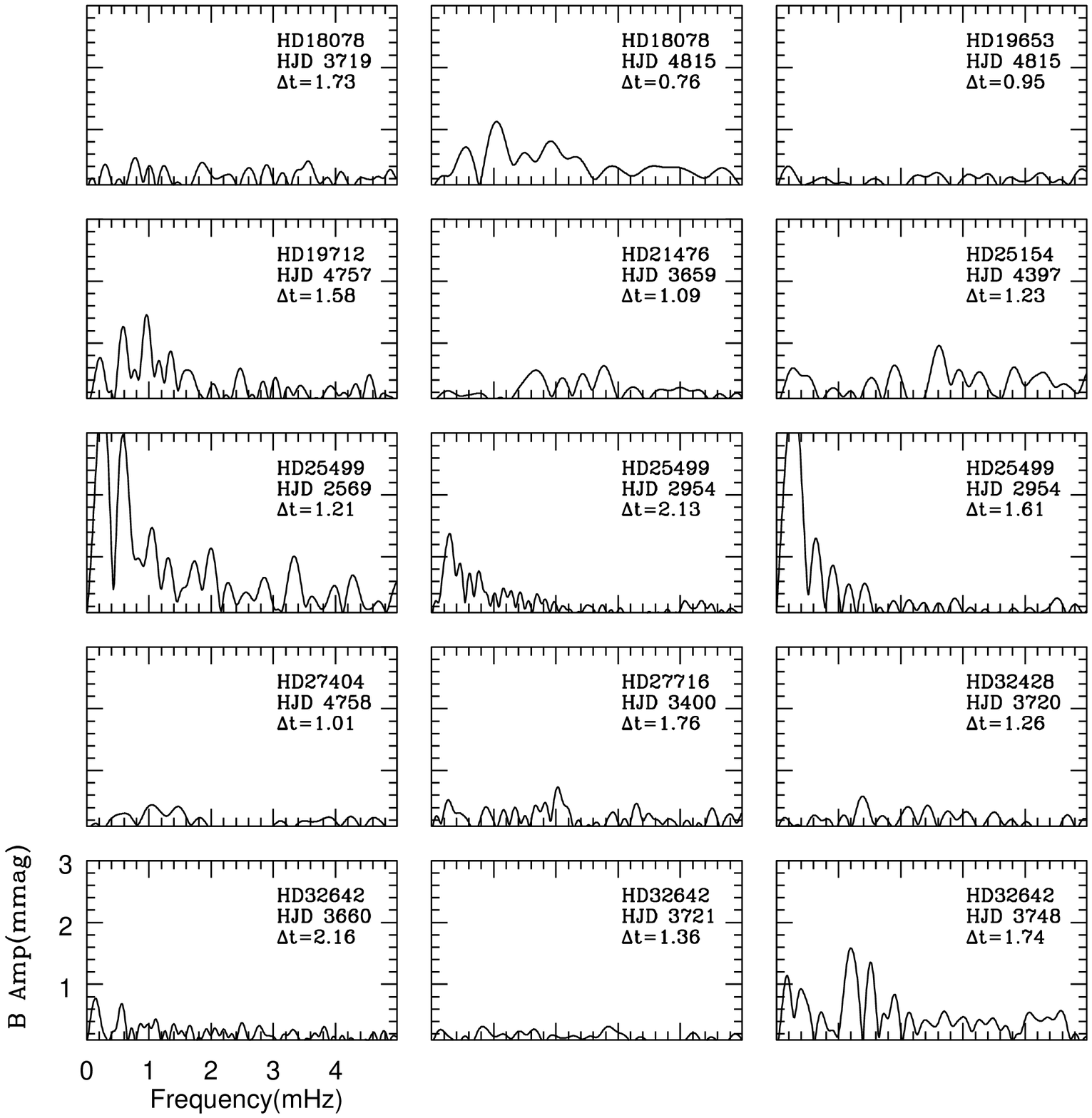}
\caption{: Cont.}
\end{figure}

\addtocounter{figure}{-1}
\begin{figure}[ht]
\centering
\includegraphics[width=16cm,height=19cm]{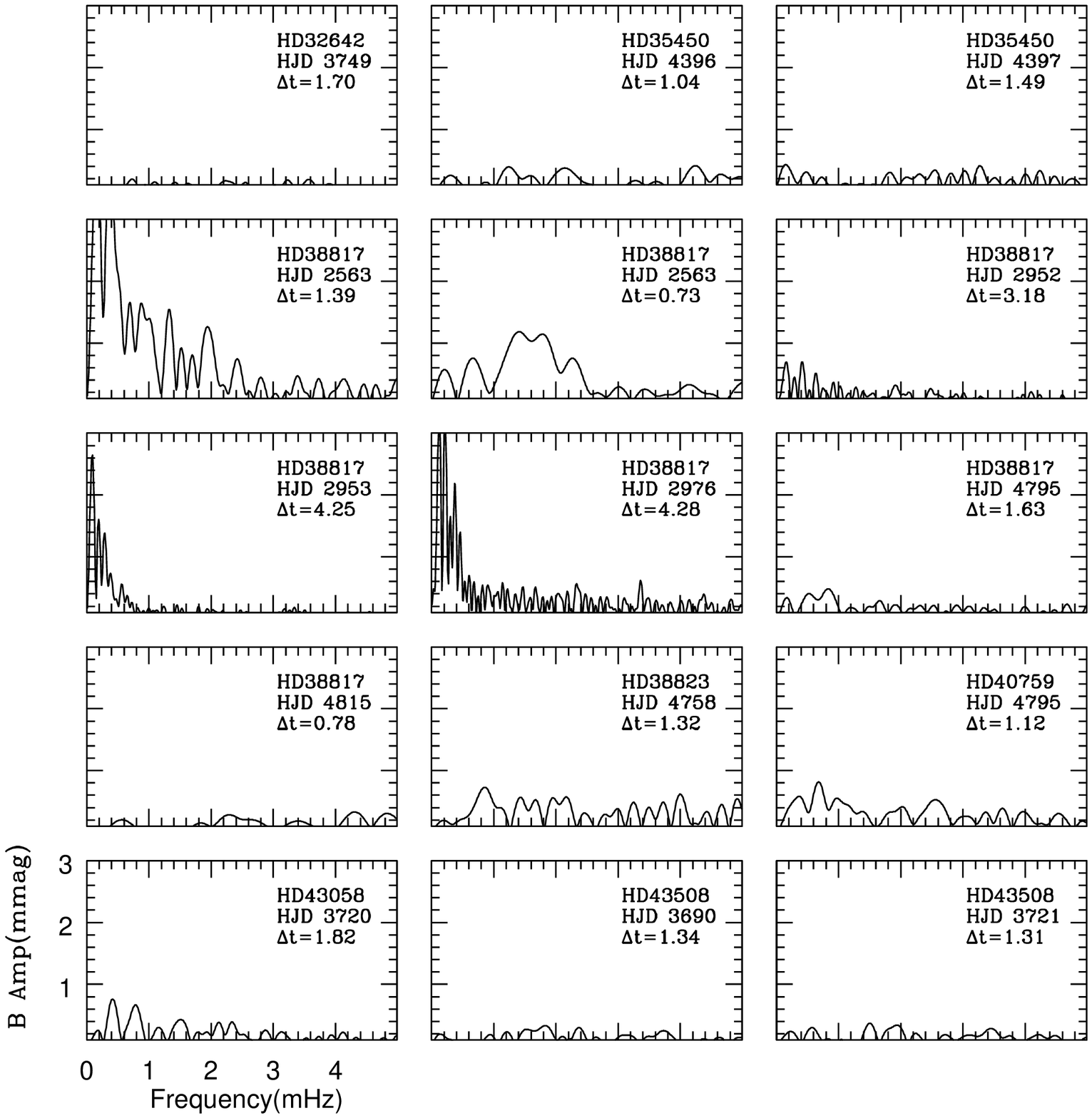}
\caption{: Cont.}
\end{figure}

\addtocounter{figure}{-1}
\begin{figure}
\centering
\includegraphics[width=16cm,height=19cm]{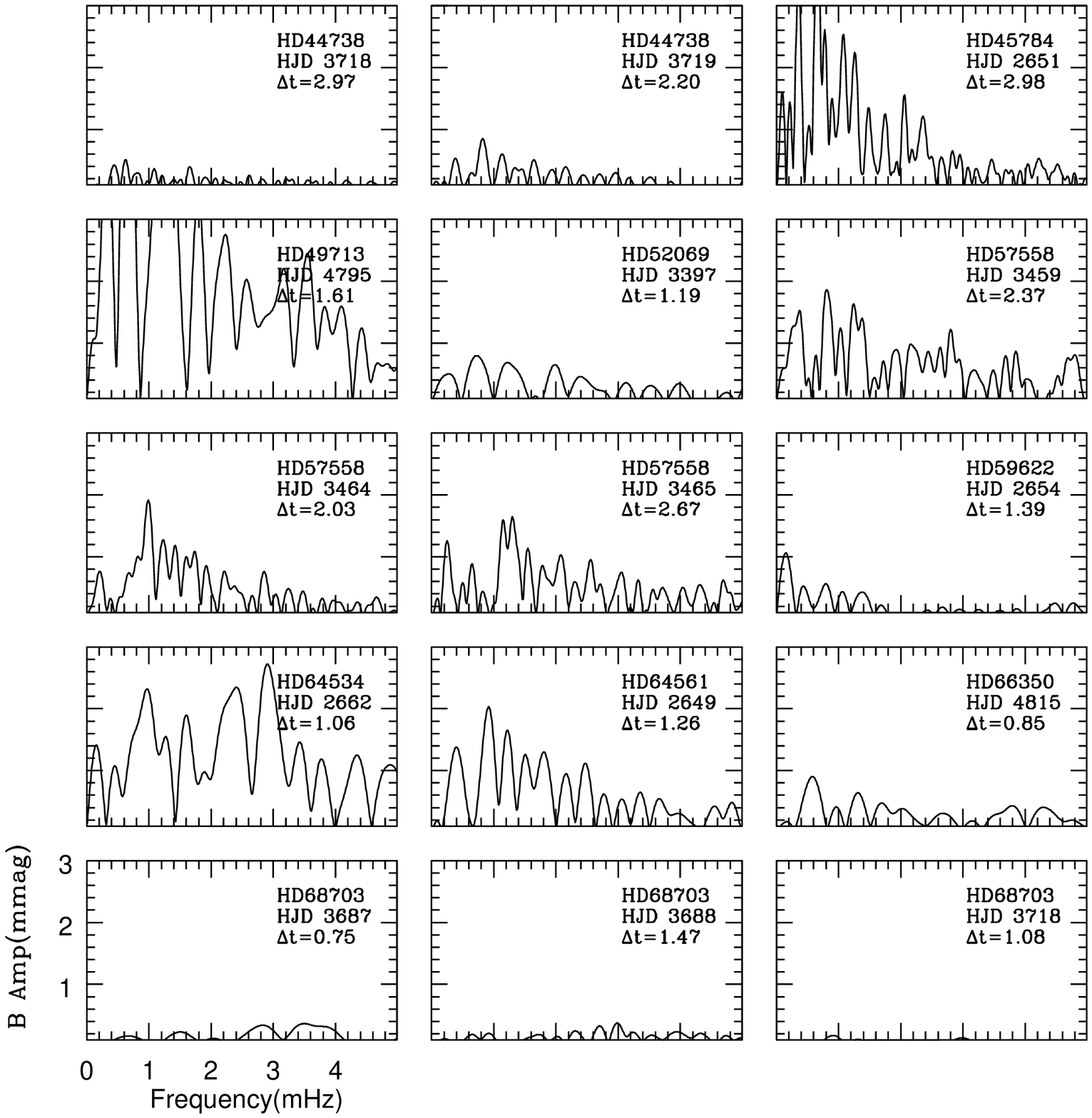}
\caption{: Cont.}
\end{figure}

\addtocounter{figure}{-1}
\begin{figure}
\centering
\includegraphics[width=16cm,height=19cm]{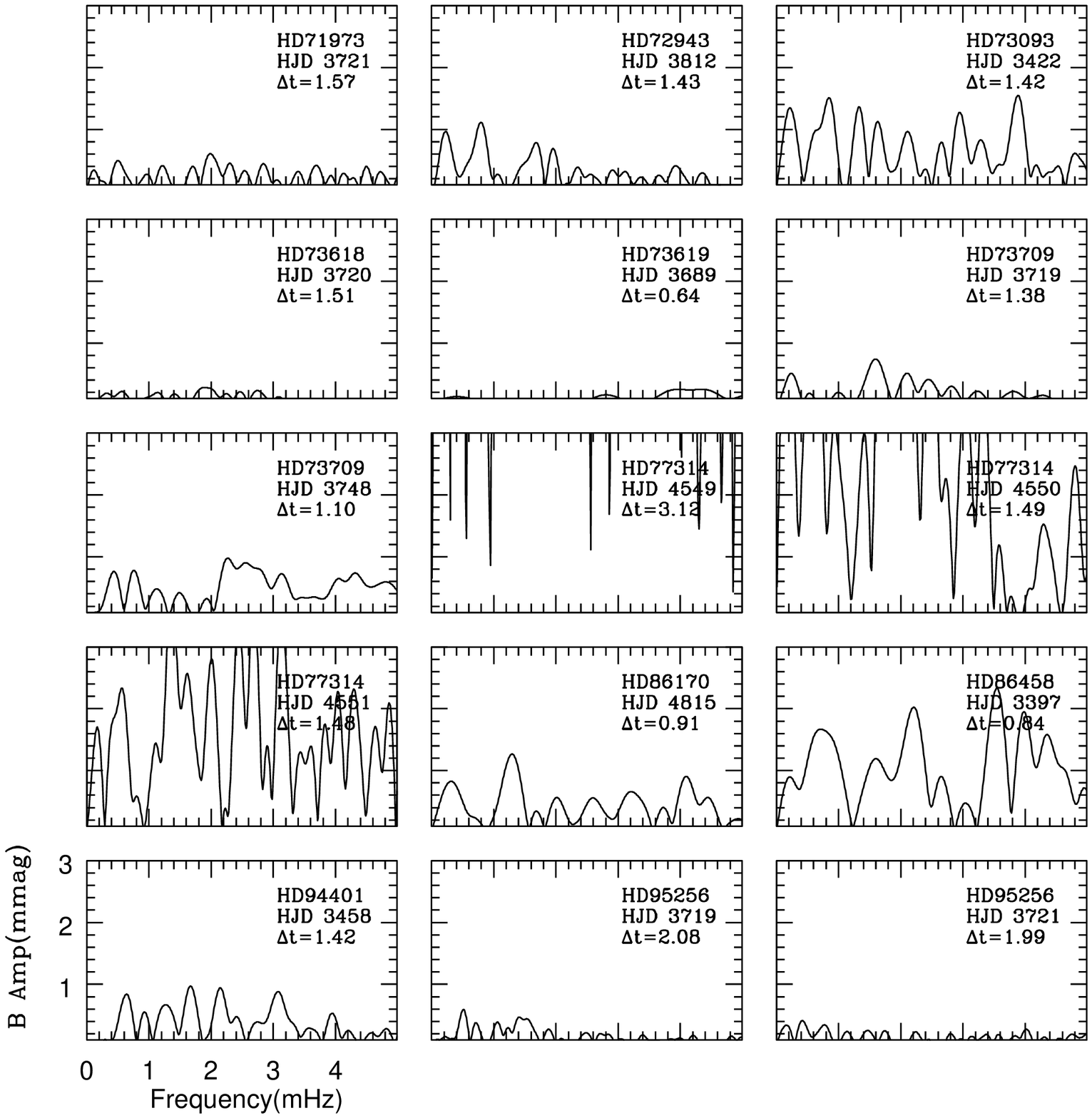}
\caption{: Cont.}
\end{figure}

\addtocounter{figure}{-1}
\begin{figure}
\centering
\includegraphics[width=16cm,height=19cm]{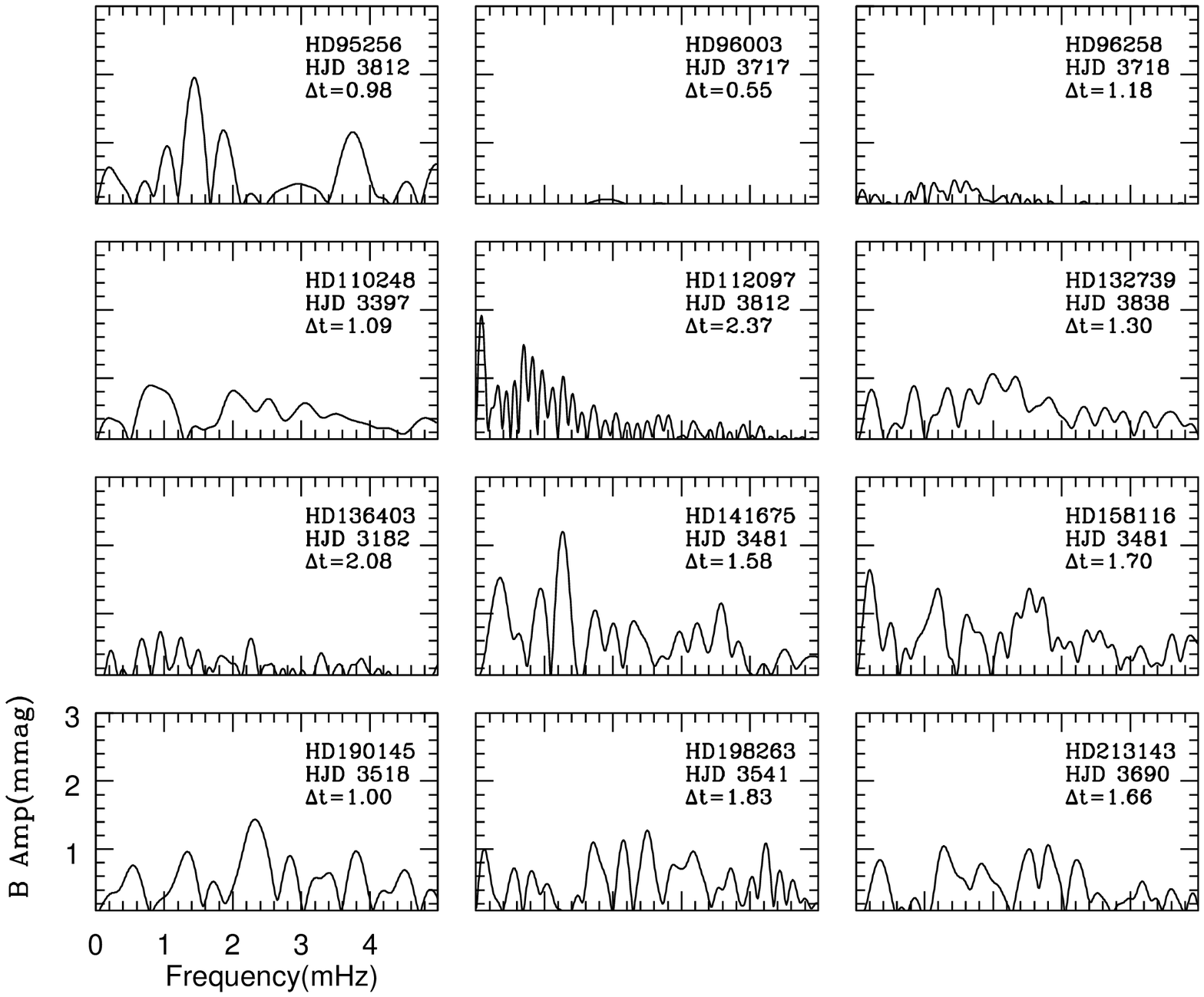}
\caption{: Cont.}
\end{figure}


\begin{thebibliography}{}
\bibitem[2004]{amado04} Amado, P. et al., 2004, in The A-Star Puzzle, eds. J. Zverko, W. W. Weiss, J. \v{Z}i\v{z}\v{n}ovsk\'y, S. J. Adelman, W. W. Weiss, Cambridge University Press, IAUS, 224, 863 \
\bibitem[2001]{balmforth01} Balmforth, N. J., Cunha, M. S., Dolez, N., Gough D. O., \& Vauclair, S., 2001, MNRAS, 323, 362\
\bibitem[2002]{bigot02}Bigot, L. \& Dziembowski, W. A., 2002, A\&A 391, 235 \
\bibitem[1970]{breger70} Breger, M., 1970, ApJ, 162, 597 \
\bibitem[2009]{bruntt09} Bruntt, H. et al., 2009, MNRAS, 396, 1189\
\bibitem[2003]{bychkov03} Bychkov, V. D., Bychkova , L. V. \& Madej, J., 2003, 407, 631 \ \
\bibitem[2007]{carrier07}Carrier, F., Eggenberger, P.,  Leyder, J.-C., Debernardi, Y., Royer, F., 2007, A\&A, 470, 1009 \
\bibitem[2005]{chaubey05}Chaubey, U. S. \& Kumar, N. S., 2005, BASI, 33, 371 \
\bibitem[1993]{christ93}Christensen-Dalsgaard, J., 1993, in  Inside the stars, Baglin A., Weiss W. W., eds, ASP Conf. Ser. Vol. 40, Proc. IAU Coll. 137, ASP, 483 \
\bibitem[1999]{cox99}Cox, N., Allen's Astrophysical Quantities, 1999 \
\bibitem[1975]{crawford75}Crawford, D. L., 1975, AJ, 80, 955 \
\bibitem[1979]{crawford79}Crawford, D. L., 1979, AJ, 84, 1858 \
\bibitem[2002]{cunha02} Cunha, M. S., 2002, MNRAS, 333, 47\
\bibitem[1975]{deeming75}Deeming, T. J., 1975, Ap\&SS, 36, 137  \
\bibitem[2005]{dorokhova05}Dorokhova, T. \& Dorokhov, N., 2005, JApA, 26, 223\
\bibitem[1973]{eggleton73}Eggleton, P.  P., Faulkner, J., Flannery, B.  P.,  1973, A\&A, 23, 325 \
\bibitem[2005]{elkin05}Elkin, V. G., Riley, J. D., Cunha, M. S., et al., 2005, MNRAS, 358, 665 \
\bibitem[2008a]{freyhammer08a}Freyhammer, L. M., Elkin, V. G., Kurtz, D. W., 2008a, MNRAS, 390, 257\
\bibitem[2008b]{freyhammer08b}Freyhammer, L. M., Kurtz, D. W. et al., 2008b, MNRAS, 385, 1402\
\bibitem[1996]{flower96} Flower, B. J., 1996, ApJ, 469, 355\
\bibitem[2001]{girish01}Girish, V., Seetha, S., Martinez, P., Joshi, S., et al., 2001, A\&A, 380, 142 \
\bibitem[2008]{gonz08}Gonz\'alez, J. F, Hubrig, S., Kurtz, D. W., Elkin, V., Savanov, I., 2008, MNRAS, 384, 1140 \
\bibitem[1999]{handler99}Handler, G. \& Paunzen, E., 1999, A\&AS, 135, 57 \
\bibitem[2004]{handler04}Handler, G., 2004, Communication in Asteroseismology, 145, 71 \
\bibitem[2004]{hatzes04}Hatzes, A. P. \& Mkrtichian, D. E., 2004, MNRAS, 351, 2 \
\bibitem[1998]{hauck98} Hauck, B. \& Mermilliod M., 1998, A\&AS, 129, 431 \
\bibitem[2008]{hekker08}Hekker, S. Fremat, Y.,  Lampens, P. , De Cat, P., 2008, 157, 317 \
\bibitem[2005]{henry05}Gregory, H. W.,  Francis,  F., C., 2005, AJ, 129, 202 \
\bibitem[2000]{hog00}H$\o$g, E., Fabricius, C, Makarov V. V. et al. , 2000, A\&A, 355, L27 \
\bibitem[1999]{houk99} Houk, Nancy \& Swift, Carrie, 1999, Michigan catalogue of two-dimensional spectral types for the HD Stars ; vol. 5 \
\bibitem[2003]{joshi03}Joshi, S., Girish, V., Sagar, R., Kurtz, D. W., et al., 2003, MNRAS, 344, 431  \
\bibitem[2006]{joshi06}Joshi, S., Mary, D. L., Martinez, P., Kurtz, D. W., et al., 2006, A\&A, 455, 303\
\bibitem[2006]{king06}King, H., Matthews, J. M. et al. 2007, Comm. in Asteroseismology, 2006, 148, 28\
\bibitem[2001]{kochukhov01}Kochukhov, O. \& Ryabchikova, T., 2001, A\&A, 374, 615 \
\bibitem[2004a]{kochukhov04a}Kochukhov, O., Bagnulo, S., Wade, G. A. et al., 2004a, A\&A, 414, 613 \
\bibitem[2004b]{kochukhov04b}Kochukhov, O., Drake, N. A.  et al., 2004b, A\&A, 424, 935 \
\bibitem[2006]{kochukhov06}Kochukhov, O. \& Bagnulo, S., 2006, A\&A, 450, 763 \
\bibitem[2008a]{kochukhov08a}Kochukhov, O.,  In the Proceedings of the Wroctaw HELAS, Workshop 2008, M. Breger, W. Dziembowski M. Thompson eds., Communications in Asteroseismology, 2008a, 157, \
\bibitem[2008b]{kochukhov08b}Kochukhov, O., Ryabchikova, T., Bagnulo, S., LoCurto, G., 2008b, A\&A, 479, L29 \
\bibitem[1985]{kreidl85}Kreidl, Tobias J., 1985, IBVS, 2739 \
\bibitem[2006]{kudryavtsev06}Kudryavtsev, D. O., Romanyuk, I. I., Elkin, V. G., Paunzen, E., 2006, MNRAS, 372, 1804.\
\bibitem[1978]{kurtz78}Kurtz, D. W., 1978, IBVS, 1436, 1 \
\bibitem[1982]{kurtz82}Kurtz, D. W., 1982, MNRAS, 200, 807 \
\bibitem[1989]{kurtz89}Kurtz, D. W., 1989, MNRAS, 238, 1077 \
\bibitem[2000]{kurtz00}Kurtz, D. W. \& Martinez P., 2000, 9, 253\
\bibitem[2005]{kurtz05}Kurtz, D. W., Cameron, C., Cunha, M. S., et al. 2005, MNRAS, 358, 651 \
\bibitem[2006a]{kurtz06a}Kurtz, D. W., Elkin, V. G., Mathys, G., 2006a, MNRAS, 370, 1274 \
\bibitem[2006b]{kurtz06b}Kurtz, D. W., Elkin, V. G., Cunha, M. S. et al. 2006b, MNRAS, 372, 286 \
\bibitem[2008]{leblanc08}Leblanc, F \& Alecian, G., 2008, A\&A,477, 243\
\bibitem[2009]{leblanc09}Leblanc, F., Monin, D. Hui-Bon-Hoa, A., Hauschildt, P. H., 2009, A\&A, 495, 937 \
\bibitem[2003]{lueftinger03}Lueftinger, T., Kuschnig, R., Piskunov, N. E., Weiss, W. W., 2003, A\&A, 406,
1033\
\bibitem[1983]{michaud83}Michaud, G., Tarasick, D., Charland, Y., Pelletier, C., 1983, ApJ, 269, 239 \
\bibitem[2003]{mkrtichian03}Mkrtichian, D. E., Hatzes, A. P., Kanaan, A., 2003, MNRAS, 345, 781\
\bibitem[1994]{martinez94}Martinez, P., \& Kurtz, D. W., 1994, MNRAS, 271, 129 \
\bibitem[1999]{martinez99}Martinez, P., Kurtz, D. W. et al., 1999, MNRAS, 309, 871 \
\bibitem[2001]{martinez01}Martinez, P., Kurtz, D. W., Ashoka, B. N., Chaubey, U. S., Girish, V., Joshi, S., et al. 2001, A\&A, 371, 1048\
\bibitem[2006]{mary06}Mary, D. L., 2006, A\&A, 452, 715 \
\bibitem[1970]{michaud70}Michaud, G. 1970, ApJ, 160, 641 \
\bibitem[1981]{michaud81}Michaud, G., Charland, Y., Megessier, C. 1981, A\&A, 103, 244 \
\bibitem[1985]{moon85}Moon, T. \& Dworetsky, M. M., 1985, MNRAS, 217, 305 \
\bibitem[1993]{nelson93}Nelson, M. J., \& Kreidl, T. J., 1993, AJ, 105, 1903 \
\bibitem[1991]{renson91}Renson P., Gerbaldi M., Catalano F. A., 1991, A\&AS, 89, 429\
\bibitem[2002]{ryabchikova02}Ryabchikova, T., Piskunov, N., Kochukhov, O., et al. 2002, A\&A, 384, 545\
\bibitem[2005]{ryabchikova05}Ryabchikova, T., Leone, F., Kochukhov, O. 2005, A\&A, 438, 973\
\bibitem[2006]{ryabchikova06}Ryabchikova, T., Ryabtsev, A., Kochukhov, O., Bagnulo, S. 2006, A\&A, 456,
329\
\bibitem[2007]{ryabchikova07}Ryabchikova, T., Sachkov, M., Weiss, W. W. et al., 2007, A\&A, 462, 1103\
\bibitem[2005]{saio05}Saio, H., 2005, MNRAS, 360, 1022 \
\bibitem[2000]{sagar00} Sagar, R., Stalin, C. S., Pandey, A. K., et al. 2000, A\&AS, 144, 349 \
\bibitem[2001]{stalin01}Stalin, C. S., Sagar, R., Pant, P., et al., 2001, BASI, 2001, 29, 39 \
\bibitem[2000]{turcotte00}Turcotte, S., Richer, J., Michaud, G., Christensen-Dalsgaard, J. 2000, A\&A, 360, 603 \
\bibitem[2007]{Lwn07} van Leeuwen, F, 2007, \aap, 474, 653 \
\bibitem[2004]{vauclair04}Vauclair, S. \& Th\'ado, S., 2004, A\&A, 425, 179\
\end{thebibliography}
\end{document}